\documentclass [11pt,letterpaper]{JHEP3}
\usepackage{epsfig,array,cite,latexsym,amsmath,oldgerm}
\usepackage{graphicx}
\usepackage{epsfig}
\usepackage{epic}       
\usepackage{cite}	
\usepackage{psfrag}     
\usepackage{bm}		
\usepackage{verbatim}   
\advance\parskip 1.0pt plus 1.0pt minus 2.0pt	
\def\coeff#1#2{{\textstyle {\frac {#1}{#2}}}}
\def\half{\coeff 12}

\def\d{{\rm d}}
\def\tr{{\rm tr}}

\def\beq{\begin{equation}}
\def\eeq{\end{equation}}
\newcommand\bal{ \begin{align}}
\newcommand\eal{\end{align} }

\newcommand\eqn[1]{\label{eq:#1}} 
\newcommand\Eq[1]{Eq.~\eqref{eq:#1}} 

\newcommand\bfs{\mathbf{S}}
\newcommand\bfz{\mathbf{Z}}
\newcommand\bfl{\boldsymbol{\Lambda}}
\newcommand\bfXi{\boldsymbol{\Xi}}

\newcommand{\CD}{{\cal D}}

\newcommand{\CN}{{\cal N}}

\newcommand{\CQ}{{\cal Q}}

\newcommand{\rma}{{\textswab{a}}}
\newcommand{\bfn}{{\bf n}}

\newcommand{\ah}{\mathbf{\hat{e}}_a}
\newcommand{\bh}{\mathbf{\hat{e}}_b}

\DeclareMathOperator{\Tr}{Tr\,}

\newcommand{\sla}[1]%
        {\kern .25em\raise.18ex\hbox{$/$}\kern-.55em #1}

\newcommand{\mybar}[1]%
        {\kern 0.6pt\overline{\kern -0.6pt#1\kern -0.6pt}\kern 0.6pt}
\newcommand{\dig}{\kern-1.5pt \raisebox{.9ex}{$\cdot$}  \kern1.5pt
  \raisebox{0ex}{${\mathbf\cdot}$}\kern1.5pt \raisebox{-.9ex}{$\cdot$}} 
\newcommand{\digb}{\kern-1.5pt \raisebox{.75ex}{$\cdot$}  \kern1.5pt
  \raisebox{0ex}{${\mathbf\cdot}$}\kern1.5pt \raisebox{-.75ex}{$\cdot$}} 
\newcommand{\digc}{\kern-1.5pt \raisebox{1.05ex}{$\cdot$}  \kern1.5pt
  \raisebox{0ex}{${\mathbf\cdot}$}\kern1.5pt \raisebox{-1.05ex}{$\cdot$}} 
\preprint {INT-PUB 04-21}
\title
    {%
    Regularization of Non-commutative SYM  by Orbifolds with Discrete Torsion
and $SL(2, \mathbb Z)$ Duality
    }%
\author
    {%
     Mithat \"Unsal
    \\Department of Physics
    \\University of Washington
    \\Seattle, Washington 98195--1560
    \\Email: 
    \parbox[t]{2in}{\email {mithat@phys.washington.edu} }
    }%
\abstract
    {%
We construct  a nonperturbative regularization for Euclidean noncommutative
supersymmetric  Yang-Mills theories  with four ($\CN= (2,2)$) , 
eight ($\CN= (4,4)$)  and sixteen ($\CN= (8,8)$) supercharges 
in two dimensions.  The 
construction relies on orbifolds with discrete torsion, which  allows     
noncommuting space dimensions to be generated  dynamically from zero 
dimensional matrix model in the deconstruction limit. We also 
nonperturbatively prove  that the twisted  topological sectors of 
ordinary supersymmetric Yang-Mills theory     
are equivalent  to a noncommutative field  theory on the topologically 
trivial sector with reduced rank and quantized noncommutativity parameter.
The key point of the proof is 
to reinterpret 
't Hooft's twisted boundary condition as an orbifold with discrete 
torsion by lifting the lattice theory to a zero dimensional matrix theory. 
  }%


\begin {document}
\setlength{\baselineskip}{1.25\baselineskip}
\section {Introduction}

Supersymmetric Yang-Mills  theories on  noncommutative space-time 
geometries appeared in  
string theory via the  toroidal compactification of 
Matrix theory \cite{Connes:1997cr, Douglas:1997fm} with a certain background 
flux, as well as  through quantization of    open strings ending on 
D branes  in the presence of  B-fields along the world volume of D brane  
\cite{Seiberg:1999vs,  Ardalan:1998ce}.
 From a field theoretic 
perspective,  field   theories on a noncommutative space \cite{Connes:1987ue,
Filk:1996dm}
are interesting  as a new universality class. 
Even though these theories are nonlocal
and  lack Lorentz invariance,  they are believed to be 
sensible quantum field theories because of their UV completion. 
(See reviews  \cite{Szabo:2001kg, Douglas:2001ba, Konechny:2000dp}.)  

One of the main problems to understand is  the existence of the 
quantized noncommutative field theory.
To answer this question requires a nonperturbative definition.
The renormalization group properties 
of noncommutative field theories may
be used with a nonperturbative 
regulator  to address the existence problem.   
Proper understanding of the renormalization group flow for these theories 
is also a fundamental 
issue.   As noncommutative  theories show interesting UV and IR  
mixing, their renormalization group properties might 
in principle be quite different from  local quantum field theories
\cite{ Sheikh-Jabbari:1999iw, Minwalla:1999px}. 

Solid progress  have  been made by the proposal of a 
 constructive definition 
of  noncommutative pure Yang-Mills theory 
\cite{Ambjorn:1999ts, Ambjorn:2000nb, Ambjorn:2000cs, Griguolo:2003kq}.  
Remarkable 
properties of these theories such as  Morita equivalence 
(T duality),  UV/IR mixing characteristic have been shown nonperturbatively
\cite{ Ambjorn:2000nb, Ambjorn:2000cs}. As a special case, 
 it has  been shown that the  twisted Eguchi-Kawai model 
\cite{Eguchi:1982ta,  Gonzalez-Arroyo:1982hz} is Morita equivalent 
to a noncommutative lattice gauge theory \cite{Ambjorn:2000nb, 
Griguolo:2003kq }. 

However, the noncommutative gauge theories emerging as limits of 
string theory are 
generally extended supersymmetric Yang-Mills theories.  
For example, 
the SYM theory on a cluster of D branes is the maximally supersymmetric 
gauge theory with sixteen supersymmetries. 
Even without incorporating noncommutativity, a nonperturbative 
definition of supersymmetric theories is  a long standing problem. Recently,  
a lattice regularization of  ordinary 
SYM  theories was presented  
 for both spatial \cite{Kaplan:2002wv} 
 and Euclidean lattices \cite{Cohen:2003xe, Cohen:2003qw, Kaplan:2005ta}. 
The supersymmetric lattices are obtained by an orbifold projection of 
a supersymmetric parent matrix  theory that has as much supersymmetry as the
target theory.  The projection leaves a subset of the supersymmetries 
(as few as one exact supersymmetry) realized 
exactly on lattice,  protecting  the theory from unwanted relevant operators 
in the continuum limit.  The orbifold projection technique  
had been introduced  in the field theory language in  \cite{Douglas:1996sw} 
and the supersymmetric lattice constructions 
are motivated by deconstruction of supersymmetric gauge theories 
\cite{Arkani-Hamed:2001ca,Arkani-Hamed:2001ie} 
\footnote{An alternative approach to lattice supersymmetry   based on 
twisted supersymmetry algebra  is  given in 
\cite{Catterall:2003wd, Catterall:2004np}, \cite{Sugino:2004uv}, 
\cite{D'Adda:2004jb} and references therein.}.       
The construction for supersymmetric spatial lattices had been  extended 
to incorporate noncommutative 
spatial lattices in \cite{Nishimura:2003tf}. In particular, they construct 
$\CN=8$ noncommutative SYM theory in $d=3$ by generalizing the results in 
\cite{Kaplan:2002wv}.  Also see \cite{Adams:2001ne, Dorey:2003pp}. 

The primary  purpose  of this  work is   to  construct a full non-perturbative 
regulator for  noncommutative Euclidean  
extended SYM theories  $\CN= (2,2)$, $\CN= (4,4)$  and $\CN= (8,8)$ in $d=2$  
dimensions with gauge group $U(k)$. These theories are the dimensional 
reduction
 of $\CN=1$ supersymmetric 
Yang-Mills theories from  $d=4,6,10$ dimensions respectively  
 to $d=2$ dimensions.   
Since we are concerned about Euclidean theories, in  principle, 
we evade the questions associated 
with timelike noncommutativity \cite{Seiberg:2000ms, Gopakumar:2000na} 
henceforth it will not  be discussed here. 

We use orbifold projection  with discrete torsion  
\cite{Vafa:1986wx, Vafa:1994rv, Douglas:1998xa, Douglas:1999hq, 
Sharpe:2000ki}  
to generate the regulated 
noncommutative SYM theory.  
But a naive implementation generates a noncommutative 
moduli space which does not lead to a sensible continuum limit, i.e the 
deconstruction limit does not generate the target theory with a 
noncommutative base space.   
In order to obtain  a sensible continuum, it is necessary to start with a 
variant of the parent theory, which can be thought of as a deformation of the
``superpotential'' (in the context of spatial noncommutative supersymmetric 
lattices, this has been suggested in 
\cite{Nishimura:2003tf}.)  
The key issue here  is that  orbifold theory is a priori 
 a matrix theory 
which has the topology of a discrete torus  but no dimensionful parameter 
to be identified as a lattice spacing. The vacuum expectation value of the 
moduli fields determine the inverse lattice spacing under appropriate 
circumstances, 
hence far  from the origin of moduli space corresponds to taking the 
lattice spacing to zero. Without the deformation, one of the moduli fields 
is frozen at  the origin and hence there is no lattice interpretation.

The deformed parent matrix theory has much less supersymmetry  than  
its standard   counterpart.  
But orbifold projection is arranged in such a way that the daughter theory  
respects the same supersymmetry of the deformed  theory. In other words, 
if the parent theory  were not deformed, 
the orbifold projection 
would break all the supersymmetries but  the ones  which are compatible 
with deformed parent theory.  The  standard  
 parent matrix theories possess 
$\CQ=4,8,16$ supersymmetries, whereas their   deformed counterparts  have     
$\CQ=1,2,4$ exact supersymmetries which is also the amount of supersymmetry 
inherited by the resulting  $d=2$ dimensional  lattices. These lattices
support the  $\CN= (2,2)$, $\CN= (4,4)$  and $\CN= (8,8)$ noncommutative 
SYM theories, respectively  in their moduli space. 
 
The orbifold projection with discrete torsion of the deformed  parent matrix 
theory is the lattice regularization  of the noncommutative SYM theory on 
a discrete torus.  The complete set of solutions of orbifold constraints form 
the basis for the noncommutative lattice. 
The algebra of (matrix valued) functions on the noncommutative discrete 
torus is in fact the same as the algebra of  
ordinary (matrix valued) functions on the commutative discrete 
torus where  the usual (matrix)  
product is modified to the lattice Moyal $\star$ product. This construction 
will basically form a full nonperturbative regulator  for Euclidean 
noncommutative SYM theories in two dimensions. 
  
The second main objective  of our work is to show that the nontrivial 
topological 
sectors of ordinary extended SYM theories are equivalent to noncommutative 
SYM theories with reduced rank at a nonperturbative level. To achieve this 
goal,  
we consider  ordinary  extended SYM theories 
with  background nonabelian  magnetic fluxes, which correspond to 
 the nontrivial topological sectors of the theory \cite{'t Hooft:1979uj, 
't Hooft:1977hy}. Since the matter fields (both fermions and bosons)
 in  all of our target theories are in adjoint representation, and hence 
are invariant 
under the center of the gauge group,  we have a  $ (U(1)\times SU(k))/{\mathbb Z_k}$ theory. 
This in turn allows us to choose different classes of boundary conditions for 
the superfields, without harming  the exact supersymmetry \cite{Witten:1982df}
of the lattice. 
  By solving twisted boundary conditions,   we nonperturbatively 
prove that an ordinary  rank $k$ 
extended SYM theory with a  definite topological charge $q$   is equivalent 
to a {\it noncommutative} SYM theory with a reduced rank $k_0$ and  
periodic boundary conditions  on a larger size torus. 
The pivotal point of the proof is realizing that  { \it 
't Hooft's twisted boundary  
condition is  in fact an  orbifold with discrete torsion when the lattice 
gauge theory is expressed as a  zero dimensional  matrix theory.} 
More interestingly, the discrete torsion (under the given circumstances) turns
out to be topological and is given in terms of the topological numbers of the 
ordinary SYM  theory as 
the ratio of integer flux $q$ to the rank $k$ of gauge group or 
equivalently, the ratio of 
first Chern character to zeroth Chern character.   

 The organization of the paper is as follows: 
In section two, the technique for obtaining a noncommutative lattice 
from orbifold with discrete torsion is reviewed. 
In section three, we  construct 
a lattice action for the noncommutative $\CN=(2,2)$  theory in $d=2$
 dimensions in detail. We show that the continuum limit with finite volume 
is compatible  
with a finite noncommutativity parameter, and this yields the 
 $\CN=(2,2)$ theory in two 
dimensions. In section four, we prove the $SL(2,\mathbb Z)$ duality 
(which is part of T duality)  nonperturbatively. 
In section five and six, we give the results for   noncommutative 
$\CN=(4,4)$ and  $\CN=(8,8)$ theories without too much detail. 
Then, we discuss some  implications of results.
\section{Non-commutative lattice from orbifold with discrete torsion}
\label{sec:nclattice}
In this section, we  first solve the orbifold conditions 
and  obtain a complete and orthonormal   basis for the noncommutative 
two dimensional lattice.  Generalization to higher dimensions is 
straightforward.
 This construction will allow us 
to write  a supersymmetric lattice gauge theory as a zero dimensional 
matrix theory in terms of matrices satisfying the orbifold constraints.  
Thus, we will be able to maneuver  from matrix to lattice or 
vice versa whenever  needed. 
Then by simple manipulations,  we show that the algebra of
 functions on the noncommutative discrete torus 
is the same as the algebra 
of ordinary  functions on the commutative discrete torus where 
the usual  product is deformed to the lattice Moyal $\star$ product. 
Most of the  results of this section are known in the literature  
\cite{Bars:1999av, Ambjorn:1999ts,  Nishimura:2003tf}. The results which 
will be used in subsequent sections are presented here for the clarity of 
discussion. 

We start with $d=0$  dimensional  parent matrix theory with a $U(N_c)$ 
gauge group, where $N_c = m_1q m_2qk$ is the number of colors and 
 $ m_1, m_2 , q , k$ are arbitrary integers. The integer $k$ will serve 
as the rank of the gauge group of the target theory, $(m_1q)^2$ will be the 
number of sites on the two dimensional lattice and  $m_2$  will play some 
 role in the noncommutativity parameter. 
In all of our examples, 
 the parent matrix theories are the   variations  of zero 
dimensional SYM theories with $\CQ=4,8, 16$ supersymmetries. 
The global nongauge symmetry group $G_R$ has at least  a  $U(1)^2$ subgroup 
which 
will be sufficient for our purpose. Any field in the parent matrix theory  
carries 
 charges $(r_1, r_2)$ under the $U(1)^2$ which determines how they  
get projected. However, this section is not specific to supersymmetric 
theories. Any gauge field theory with adjoint matter on a  noncommutative 
or commutative lattice can be obtained 
in the same way. Hence, we will try to minimize discussions related to 
supersymmetry  until subsequent sections.    

Let $\Phi$ be a generic  matrix in the parent matrix theory and 
let us consider 
a $\mathbb Z_{m_2q} \times \mathbb Z_{m_2q}$ orbifold projection.    
The sub-blocks of the matrix $\Phi$ remaining invariant under the  
constraints 
\begin{equation} 
\Phi = \omega_{m_2 q}^{r_a} \Omega_a^{\dagger} \Phi \Omega_a  \qquad a= 1, 2
\eqn{orbifoldconstraint}
\end{equation}
are said to satisfy the orbifold constraint 
and they form the field content of the orbifold theory. 
The phase  $ \omega_{m_2 q}= e^{2 \pi i/(m_2q)}$  
is $m_2q$-th primitive root 
of unity.  The $r_a$ are integer 
charges under an abelian subgroup of the global  symmetry group $G_R$. 
The   orbifold matrices $\Omega_a \in   U(N_c)$ are rank-$N_c$
matrices forming a  projective representation of the group $\mathbb Z_{m_2q}$.
A particular matrix representation for  $\Omega_a$ may be written as    
\begin{eqnarray}
 \Omega_1 &=& (U_{m_1m_2q})^{m_1} \otimes ({V_q^{\dagger}})^{p}
\otimes {\bf 1}_k 
\nonumber \\
\Omega_2 &=& (V_{m_1m_2q})^{m_1} \otimes U_{q}^{\dagger} \otimes {\bf 1}_k
\eqn{orbifoldmatrices}
\end{eqnarray} 
where the subscripts of $U, V,{\bf 1}$  denote their rank  
and superscripts stand for  the corresponding  power.   
The  
 $U_q$   and $V_q$  are called clock (position) and shift (translation)
 matrices and are given by
\begin{equation} 
U_{q} = \left( \begin{array}{cccc}
               1 & &     &         \\
                 & \omega &&          \\ 
                 && \ddots &       \\
               &&& \omega^{q-1}
\end{array}
\right) 
\qquad 
V_{q} = \left( \begin{array}{ccccc}
               0 & 1 &&&   \\
                & 0 & 1 &&    \\
               & &  \ddots & \ddots &          \\ 
                & & & 0  & 1      \\
               1 &&&& 0
\end{array}
\right) 
\eqn{UV}
\end{equation}
and they satisfy the algebra  
\begin{equation}
U_qV_q= \omega_q^{-1}V_qU_q \, ,  \qquad 
\end{equation}
We choose $p$ and $q$ as coprime for later convenience.

The orbifold matrices $\Omega_a$ may be thought as generators of 
the projective   representation of the abelian group $\mathbb Z_{m_2q}\times 
\mathbb Z_{m_2q}$; each factor    is embedded in $U(N_c)$.
Even though the  group is abelian, the two sets of generators 
$\Omega_a$'s in  \Eq{orbifoldmatrices} do not commute with each other. 
Instead they satisfy     
\begin{equation}
\Omega_1\Omega_2 = e^{2 \pi i \theta} \Omega_2\Omega_1
\qquad {\rm{where}} \qquad 
\theta =  -\frac{m_1}{m_2q} + \frac{p}{q}  \qquad {\rm modulo} \,\, {\mathbb Z}
\end{equation}
and the rational number  $\theta \in [0,1)$  which appears as  the 
noncommutativity of 
$\Omega_a$'s  is called    the {\it discrete torsion}. It is not  the 
noncommutativity parameter of the discrete torus on which the SYM theory 
lives, but  it turns out to 
be closely related. There is a good reason that we have split the definition 
of $\theta$ 
into two pieces. In fact, as we will explain  in detail in  section 
\ref{sec:Tduality}, 
the second term in the sum may be thought of as due to a 
background magnetic flux. 

In order to find the most general solution to the constraints 
\Eq{orbifoldconstraint}, we first introduce a set of matrices $D_a$ 
which play two important roles:  
{\it i)} They are displacement operators on the 
noncommutative lattice; {\it ii)} They transform  as if they  carry  
a definite 
${\bf r}$-charge. We will first explain the second property and use it to 
{\it neutralize}  the orbifold constraint 
\cite{Ambjorn:1999ts, Nishimura:2003tf}. \footnote{ These matrices are also 
known as twist-eaters in the literature. But here we prefer not to call them 
 that for two reasons. One is that they don't really serve 
 that purpose here, and the other reason is that the real twist 
eating solutions are associated with 't Hooft's boundary conditions and 
 will show up in section \ref{sec:Tduality}}
 
The first property will be better appreciated 
when the real space basis for the lattice is introduced.   
In order to find solution to \Eq{orbifoldconstraint},  let  us introduce   
$D_1, D_2 \in U(N_c)$  as
\begin{eqnarray}
 D_1 &=& V_{m_1 m_2 q}^{\dagger} \otimes {\bf 1}_q \otimes {\bf 1}_k 
\nonumber \\
D_2 &=& U_{m_1m_2q} \otimes {\bf 1}_{q}\otimes {\bf 1}_k
\eqn{twisteaters}.
\end{eqnarray} 
The  remarkable property of these matrices is that $D_1$ transforms  as 
if it  has a charge vector  ${\bf r}= (1,0)$ and similarly, $D_2$ transforms 
like it has  a charge vector 
$(0,1)$. In general, $D_1^{r_1} D_2^{r_2}$ obeys the orbifold constraints that 
a field with charge vector 
$r_1(1,0)+ r_2(0,1)= (r_1,r_2)$ would obey. Namely, 
\begin{equation} 
(D_1^{r_1} D_2^{r_2}) = 
\omega_{m_2 q}^{ r_a} \,\, \Omega_a^{\dagger} (D_1^{r_1} D_2^{r_2}) 
 \Omega_a \qquad a=1,2.
\eqn{Dalgebra}
\end{equation}
Since $D_1$ and $D_2$ can also be interpreted  as finite translation operators 
in orthogonal  directions in noncommutative lattice, they do not commute with 
each other. Instead, they 
satisfy the relation 
\begin{equation}
D_1 D_2 = \omega_{m_1 m_2 q}^{-1} D_2 D_1 \,. 
\end{equation}   
The main significance of $D_a$'s may be realized as follows. Let $\Phi$ have 
${\bf r}$-charge vector $(r_1, r_2)$.   
 Let us define ${\widetilde \Phi}$ by the equation 
\beq
\Phi= \left\{    \begin{array}{cc}
 \widetilde \Phi D_1^{r_1}  D_2^{r_2} & \qquad  {\rm if} \,\, r_1 > 0, 
\, r_2 > 0 \\
 D_1^{r_1} \widetilde \Phi  D_2^{r_2} & \qquad  {\rm if} \,\,r_1 < 0, 
\, r_2 > 0 \\
 D_2^{r_2} D_1^{r_1}  \widetilde \Phi  & \qquad  {\rm if} \,\, r_1 < 0, 
\, r_2 < 0
\end{array} \right.
\eqn{neutralization}
\eeq
It is easy to show that ${\widetilde \Phi}$ obeys the orbifold constraint 
as if it is neutral under the global symmetry group $G_R$, i.e,   
${\widetilde \Phi}$  has charge vector ${\bf r}=(0,0)$. In that sense, 
\Eq{neutralization} is a neutralization procedure. By using 
 \Eq{orbifoldconstraint}  and  \Eq{neutralization}, we  
obtain  
%
%
the  neutral  orbifold constraints for $\widetilde \Phi$ as  
\begin{equation} 
{\widetilde \Phi} = \Omega_a^{\dagger} {\widetilde \Phi} \Omega_a 
 \qquad a= 1, 2 \,.
\eqn{neutralconstraint}
\end{equation}
Once the  neutral  constraint is solved, we can 
immediately get the solution for the  orbifold constraint  
\Eq{orbifoldconstraint} by using the $D_a$ matrices as in   
\Eq{neutralization} which follows easily from the 
invertibility of $D_a$ matrices.
 
\noindent{\it Noncommutative lattice basis and crossed product algebra }
 
To identify the set of solutions to  neutral orbifold  constraint 
\Eq{neutralconstraint}
as the noncommutative  lattice, 
we need to  construct the explicit isomorphism between the lattice 
and the matrices satisfying \Eq{neutralconstraint}. 
The complete set of commutants of the orbifold matrices can be obtained by 
using the  matrices $Z_a \in U(m_1 q m_2 q)$ 
\beq
Z_1= (U_{m_1m_2q})^{m_2} \otimes V_q^{\dagger}, \qquad
Z_2= (V_{m_1m_2q})^{m_2} \otimes ({U_q^{\dagger}})^r  \,.
\eeq
The $Z_a$'s  are constructed such 
that $ Z_a \otimes {\bf 1}_k $   commutes with  
the orbifold matrices. 
The commutators $[Z_1 \otimes {\bf 1}_k,\Omega_a]= 
[Z_2 \otimes {\bf 1}_k ,\Omega_2]=0$ are trivially satisfied. 
The condition  
for $Z_2 \otimes {\bf 1}_k $  to commute with $\Omega_1$ requires  
$\frac{1 - rp }{q}=s$ to be an  integer. The equation (also known as a first 
order Diophantine equation) 
\beq
1=  qs + pr 
\eqn{Diophantine}
\eeq
is guaranteed to have a solution (see \cite{Dummit}, page 276) for some 
integers $(s,r)$ if the greatest common divisor of $p$ and $q$ is $1$. Since 
we assumed $p$  and $q$ are co-prime,  
by construction there exists such a pair of integers $(s,r)$.  

The base matrices $Z_a$ are cyclic with period $m_1q$, and they satisfy   
$Z_a^{m_1q}= {\bf 1}_{m_1 q  m_2 q}$. 
The $Z_1$ and $Z_2$ are  projective representations of the abelian cyclic 
group ${\mathbb Z_{m_1q}}$, and   they satisfy 
\beq 
Z_1 Z_2 = e^{-2 \pi i \theta'}Z_2 Z_1 \, ,  \qquad \theta'= 
 +\frac{m_2}{m_1q}  -\frac{r}{q}   
\eqn{theta'}
\eeq
where the  rational number   $\theta' \in [0,1)$ is   the 
dimensionless noncommutativity parameter of the lattice. The discrete torsion 
$\theta$ and noncommutativity $\theta'$ are related by an 
$SL(2, \mathbb{Z})$ transformation 
\beq
\theta'= \frac{r\theta+ s}{-q \theta + p} 
\qquad {\rm{where}} \qquad
                                         \left(
                                         \begin{array}{cc}
                                          r & s \\
                                          -q & p  
                                         \end{array}
                                         \right) \in SL(2, \mathbb{Z}) \,.
\eeq  

The space of solutions to the orbifold condition is spanned by  
 $(m_1q)^2$ matrices 
\beq
\bigl\{ Z_1^{p_1} Z_2^{p_2} | \, p_1, p_2  = 1, \ldots, m_1 q \bigr\} 
\eeq
which form a complete  and  orthogonal basis for the lattice (for a proof, see
\cite{Ambjorn:1999ts}). 
This basis   is not hermitian but can be made so by introducing 
a phase factor: 
\beq
J_{\bf p}= e^{- \pi i \theta' p_1 p_2}  Z_2^{p_2} Z_1^{p_1}
\qquad
\rm where  \qquad J_{\bf p}^\dagger = J_{-\bf p} \,.
\eqn{momentumbasis}
\eeq

The  basis  we have constructed will be identified as the momentum basis 
of the lattice. It has formally the same algebraic properties as 
the plane wave basis on a noncommutative space  which satisfies  
\beq
e^{ip_i {\hat x}^i } e^{iq_j {\hat x}^j } = e^{-\frac{i}{2}p_i\theta^{ij}q_j} 
 e^{i (p_i+q_i) {\hat x}^i }
\eeq
where $p_i$ and $q_i$ label components of momenta 
and $\theta^{ij}$ is a real-valued antisymmetric matrix. 
 Using  the defining relation \Eq{momentumbasis} for the regularized 
momentum basis for the lattice,  we obtain the corresponding regularized 
algebra as 
\beq
J_{\bf p}\, J_{\bf q} = e^{- \pi i \theta'(p_1q_2 - p_2 q_1)}  
J_{\bf p+ q}=   e^{-\pi i \theta' {\bf p}\wedge{\bf q}   }  
J_{\bf p+ q} \,. 
\eqn{crossedproduct}
\eeq
The ${\bf p}$ labels momentum on a periodic lattice,  
in a two dimensional Brillouin zone  $\mathbb{Z}_L^2$ where  the addition  
of momenta is understood to be modulo $L=m_1q$. 
The algebra in \Eq{crossedproduct} is a realization of 
a  {\it crossed product algebra}.
It is easy to see that this algebra is not commutative for generic $\theta'$, 
and it 
  involves trigonometric functions as 
the structure constants:  
\cite{Fairlie:1988qd}\footnote{ The structure constants of this algebra 
show up in the momentum space formulation of the action, in particular 
at cubic and quartic  vertices. For example, this is the case in 
\cite{Sheikh-Jabbari:
1999iw} where a momentum space formulation of continuum 
noncommutative SYM is given.  
The above   algebra is a UV  regulation of the one given in 
 \cite{Sheikh-Jabbari:1999iw}. However, in the rest of this paper, 
we will mostly work in  position space.}    
\beq
[ J_{\bf p}\,,  J_{\bf q}] =- 2 i \, 
 \sin(\pi  \theta' {\bf p}\wedge{\bf q}   )\,  J_{\bf p+ q} \,.
\eeq 
The product is  associative, i.e. 
$J_{\bf p}(J_{\bf q}J_{\bf r})= (J_{\bf p}J_{\bf q})J_{\bf r}$. 
The crossed  product algebra of lattice generators 
gives the first glimpse   of  the appearance of noncommutativity of the 
base   space
in the target  theory. We emphasize that the matrices $J_{\bf p}$
form  a projective representation for the abelian group 
 $\mathbb Z_{m_1q} \times \mathbb Z_{m_1q}$ which is in fact  the lattice.  
Each point ${\bf p}$ on the momentum space lattice is represented by a 
unique matrix  $J_{\bf p}$.  
For more on projective representation of finite groups, see 
\cite{Karpilovsky}.
%

In order to  construct a noncommutative field theory with periodic 
boundary conditions,    
we need to restrict $\theta'$.  The periodicity conditions   
$J_{{\bf p}+ L{\ah}}  = J_{{\bf p}+ L{\bh}}= J_{\bf p} $ are satisfied 
if   $\, \theta'  L p_a \, $  is an even integer for all $p_a=1,\ldots, L$. 
This holds if    
\begin{equation} 
 L \theta'  = (-r m_1 + m_2 ) \in 2 {\mathbb Z}
\eqn{mod2} \ .
\end{equation} 

Now, we are ready  to write down the 
 most general  configuration of the matrix $\widetilde \Phi$ satisfying 
the neutral orbifold constraint \Eq{neutralconstraint} :  
\beq 
{\widetilde \Phi} = \sum_{{\bf p} \in \mathbb{Z}_L^2} J_{\bf p}^{\dagger}
\otimes \Phi_{\bf p} 
\eqn{matrixlattice}
\eeq 
where $\Phi_{\bf p}$ is a rank $k$   field matrix  that 
we associate to each momenta ${\bf p}$ in the Brillouin zone. 
The basis matrices 
$J_{\bf p}$ are rank $m_1 q m_2 q$ and ${\widetilde \Phi}$ is rank 
 $m_1 q m_2 q k$.  The equation has to be understood as a map  from the 
dual momentum lattice to a matrix symbol \footnote{The 
matrix ${\widetilde \Phi}$ defined as in \Eq{matrixlattice} 
is  sometimes called the Weyl symbol or matrix symbol.}  and vice versa.  
    
 The   coordinate basis for the lattice is  the discrete 
Fourier transform  of the momentum basis 
 given by   
\beq
\Delta_{\bf n} = \frac{1}{L^2} \sum_{{\bf p} \in \mathbb{Z}_L^2} J_{\bf p} 
\, \omega^{{\bf p}.{\bf n}} \, ,  \qquad 
J_{\bf p} =  \sum_{{\bf n} \in \mathbb{Z}_L^2} 
\Delta_{\bf n} \, \omega^{-{\bf p}.{\bf n}}
\eqn{basis} \,.
\eeq
The index ${\bf n}$ spans a two dimensional real space lattice where each 
component is ranging from $1, \ldots L$. One can rewrite  
\Eq{matrixlattice} in a position space lattice as 
\beq
\widetilde \Phi = 
\sum_{{\bf n} \in \mathbb{Z}_L^2} \Delta_{\bf n} \otimes 
\Phi_{\bf n},  \qquad \Phi_{\bf n} = \Tr \left( \widetilde \Phi 
(\Delta_{\bf n} \otimes {\bf 1}_{k})
\right)
\eqn{matrixlattice2}
\eeq
where $\Phi_{\bf n}$  is a $k \times k$ field matrix 
that we associate  with each site  ${\bf n} \in  \mathbb{Z}_L^2$ 
in the position space lattice.  
The second relation follows from the  completeness and orthonormality 
of the  noncommutative basis $\Delta_{\bf n}$.  
These two relations reveal  that if we know the value of a field at each 
site ${\bf n}$ on the lattice, we can merge  them into a single 
 large matrix, the matrix symbol. Or equivalently, if we know the 
matrix symbol, we can unravel it to obtain the field matrix at each site
of the lattice.  
Here, we list certain relations satisfied by basis matrices:
\footnote{ The trace is rescaled 
by a factor $\frac{m_2}{m_1}$ so that  $\Tr {\bf 1}_{ m_1q m_2q}= 
Volume(Lattice) = L^2 $, the dimensionless volume of the lattice.  
With this convention,  the identities for basis 
matrices \Eq{identities} are 
formally the same as the conventional continuum 
ones, see \cite{Szabo:2001kg}.   
}
\begin{eqnarray}
&& \sum_{{\bf n} \in 
\mathbb{Z}_L^2} \Delta_{\bf n} = {\bf 1}_{m_1qm_2q} \,,\nonumber \\
&&
\Tr ( \Delta_{\bf n})= 1 \,, \nonumber
 \\
&& \Tr (\Delta_{\bf n} \Delta_{\bf m} ) = \delta_{{\bf n},{\bf m}}
\,,  \nonumber \\
&&  \Tr (\Delta_{\bf k} \Delta_{\bf l} \Delta_{\bf m} ) = 
\frac{1}{ L^2}  \, \omega^{-\frac{2}{L \theta'} ({\bf k-m}) 
\wedge ({\bf l-m})} \,.
\eqn{identities}
\end{eqnarray}
The phase in the last formula is associated with a nonlocal cubic vertex. 
The counterpart of these relations  in commutative case is the  same  
except for the cubic and higher order vertices. For example, 
in the commutative case, the cubic vertex is 
\begin{equation}
 \Tr (\Delta_{\bf k} \Delta_{\bf l} \Delta_{\bf m} ) = 
 \delta_{\bf k,m} \,  \delta_{\bf l,m}
\end{equation}
where the locality is manifest.  
The meaning of the operators $D_a$ in \Eq{twisteaters} is also clear 
in the coordinate basis. They  act as finite difference operators on the 
noncommutative lattice, 
\begin{equation}
D_a  ( \Delta_{\bf n} \otimes \Phi_{\bf n} ) D_a^{-1} =
\Delta_{{\bf n} - \ah} \otimes \Phi_{\bf n}  
\eqn{difference}.
\end{equation}

\noindent{\it From  noncommutative basis to Moyal $\star$ product }

Next, we show that the algebra of the matrix valued functions  
on a noncommutative discrete torus can also  be regarded as 
the usual algebra of matrices  on the 
commutative discrete torus where   the usual matrix  
product  is replaced by  the lattice Moyal $\star$ product.  The construction
here also provides an unambiguous definition of the $\star$ product on the 
lattice.
Assume two matrices ${\widetilde \Phi_1}$ and ${\widetilde \Phi_2}$
  obey the  orbifold conditions \Eq{neutralconstraint}. 
Then so does their product $\widetilde \Phi_1 \widetilde \Phi_2$, which may 
be   expressed  in the complete lattice basis as   
\beq
\widetilde \Phi_1 \widetilde \Phi_2 = \sum_{{\bf n} 
\in \mathbb{Z}_L^2} \Delta_{\bf n} \otimes 
[\Phi_{1,{\bf n}}\star \Phi_{2,{\bf n}}]
\eqn{phi1phi2}
\eeq
where   we define 
$[\Phi_{1,{\bf n}}\star \Phi_{2,{\bf n}}]=  \Tr \left( \widetilde \Phi_1 
 \widetilde \Phi_2 
(\Delta_{\bf n} \otimes {\bf 1}_{k})
\right) $.  To obtain the expression for $\star$ product, consider 
 \beq
\widetilde \Phi_1 \widetilde \Phi_2 = \sum_{{\bf n,m} \in 
\mathbb{Z}_L^2} \Delta_{\bf n} \Delta_{\bf m} \otimes 
\Phi_{1,{\bf n}}\Phi_{2,{\bf m}}
\eqn{product}
\eeq
where $\Phi_{1,{\bf n}}\Phi_{2,{\bf m}}$ is the usual matrix product at this 
stage. One can calculate  the product of basis matrices 
$\Delta_{\bf n} \Delta_{\bf m}$ by using \Eq{basis} 
and the definition of the crossed product algebra \Eq{crossedproduct} : 
\begin{eqnarray}
 \Delta_{\bf n} \Delta_{\bf m} &=&  
\frac{1}{L^2}
{\sum_{{\bf s} \in \mathbb{Z}_L^2}}^{\prime} \Delta_{\bf s} \,\,
\, \omega^{-\frac{2}{L \theta'}({\bf m-s})\wedge({\bf n-s})}\
\end{eqnarray}
where the prime on the summation sign indicates that the sum is 
over ${\bf s} \in \mathbb{Z}_L^2$ 
such that either 
\begin{equation}
({\bf m-s}) \in \frac{L \theta'}{2} 
\mathbb{Z} \times \frac{L \theta'}{2} 
\mathbb{Z} \qquad  {\rm or} \qquad 
\, 
({\bf n-s}) \in  \frac{L \theta'}{2} \mathbb{Z} \times 
\frac{L \theta'}{2} \mathbb{Z} 
\eqn{sum'}
\end{equation}
modulo periodicity of the lattice.
Substituting this result into \Eq{product} and using 
orthonormality of the basis, we obtain the  definition of the Moyal 
$\star$ product on the lattice. 
We have   
\begin{eqnarray}
[\Phi_{1,{\bf s}}\star \Phi_{2,{\bf s}}] &=& 
 \frac{1}{L^2} {\sum_{{\bf n,m} \in 
\mathbb{Z}_L^2}}^{\!\!\!\!\!\prime} \,
\Phi_{1,{\bf n}}\Phi_{2,{\bf m}}\, \omega^{-\frac{2}{L \theta'}
({\bf m-s})\wedge({\bf n-s})} 
\eqn{latticeMoyal}
\end{eqnarray}
where the product of matrices within the sum is the usual matrix product.
The prime on the double sum is  to remind us that 
one of the sums is over all the  lattice  points, whereas the other sum 
is over the sublattice  \Eq{sum'} modulo periodicity $L \times L$.    
This shows that the algebra of functions  on 
the  noncommutative  discrete torus    can be traded with a  deformed 
 algebra of functions  on a commutative discrete torus. The deformed 
product \Eq{latticeMoyal} is     
 called the Moyal $\star$ product and it is nonlocal. In particular,  for the 
 minimum nontrivial value of  $\theta'$  
\beq
\theta'_{min} = \frac{2}{L}= \frac{2}{m_1q}
\eqn{thetamin}
\eeq
 both  sums run over  all sites on the lattice  
with an oscillatory   kernel. 

As a last remark on the lattice Moyal 
 $\star$  product  \Eq{latticeMoyal}, we observe that   the $\star$ product 
of a  constant 
matrix field $M$ on the lattice with a  varying matrix field
$\Phi_{1,{\bf s}}$ is just 
the usual multiplication,  
\beq
\Phi_{1,{\bf s}}\star M= \Phi_{1,{\bf s}} M 
\eqn{const}
\eeq
  In particular, $\star$ multiplication of two 
constant matrices $M$ and $N$ is just the usual multiplication
$ M \star N = M N $. 

\section{Regularization for noncommutative 
$\CN= (2,2)$ SYM in $d=2$ }  
\label{sec:reg}
The action of the parent theory is a variant of dimensional 
reduction of $\CN=1$ supersymmetric Yang-Mills theory with $U(N_c)$ gauge 
group  from $d=4$ Euclidean  dimensions to zero dimension. The standard 
reduction of $\CN=1$ SYM possess a $\CQ=4$ supersymmetry and  
$SO(4)\times U(1)$ global R-symmetry. The  $SO(4)$ symmetry  is inherited 
from the Lorentz symmetry and $U(1)$ is the global R-symmetry of the 
Euclidean  theory prior to reduction.     

The zero dimensional matrix model is composed of bosonic and fermionic 
Lie algebra valued matrices. The bosonic matrices are $\{z_1, z_2,  
\mybar z_1,\mybar z_2, d \}$  where $z_a \,(\mybar z_a) $  are 
the complexifications of dimensional reduction of the 4-vector 
gauge potential  
and  $d$ is an auxiliary field.  
The fermionic matrices are Grassmann valued 
matrices $\{\psi_1, \psi_2, \lambda, \xi\}$  which were components  of 
two independent, complex two-component spinors in four dimensions prior 
to dimensional reduction. 

We deform the reduced matrix model by a pure phase $\zeta$,   
in a way that the action only respects 
one out of four  supersymmetries and no other. We call this parent theory 
the {\it deformed matrix model}.
The action   is 
\begin{equation}
S_{{\zeta}}=
\frac{1}{g^2}\Tr\Biggl[ \frac{1}{2} d^2
+ i d [\mybar z_a, z_a ]
+2 \bigl| [z_1, z_2]_{\zeta} \bigr|^2
+
\sqrt{2}\left(\lambda\,[\mybar z_a,\psi_a] 
  + \xi \, [ z_1,\psi_2]_{\zeta} + \xi \,
[\psi_1, z_2]_{\zeta} \right)  \Biggr]\
.
\eqn{twistedmatrix}
\end{equation}
where $[z_1 ,z_2 ]_{\zeta} \equiv ( z_1z_2 - {\zeta} z_2 z_1) $ etc.  
The deformed  matrix model action \Eq{twistedmatrix}  
has all $\CQ=4$ supersymmetries for $ \zeta = 1$ and has only $\CQ=1$ for 
$\zeta \neq 1$. For $\zeta=1$, after eliminating the auxiliary field $d$, 
the bosonic part of the action is the zero dimensional counterpart of the 
sum of electric and magnetic energies, whereas  the fermionic part is 
the interaction term inherited from the covariant derivative.  
  
The supersymmetry  that the action respects may be written as 
\begin{eqnarray}
\delta z_a &&= {\sqrt 2} i \eta \psi_a   \qquad a=1,2  \nonumber \\
\delta {\mybar z}_a &&= 0 \nonumber \\
\delta  \psi_a &&= 0 \nonumber \\
\delta \lambda &&=  \eta  d \nonumber \\
\delta \xi &&= -2 i \eta  
[{\mybar z}_2, {\mybar z}_1]_{\zeta^{*}} \nonumber \\
\delta d  &&= 0 \,.
\eqn{parentsusy}
\end{eqnarray}
where $\eta$ is a Grassmann valued parameter.
The double variation of any field by  $\delta$ is identically zero.  With 
$\delta=i \eta Q= i \eta \partial_{\theta} $, the 
transformation laws can be realized in terms of the 
following $\CQ=1$ superfields:
\begin{eqnarray} 
{\bfz}_a &&=  z_a + {\sqrt 2} \theta \psi_a  \nonumber \\
\mybar z_a &&= \mybar z_a \nonumber \\
{\bfl}  && =  \lambda -i \theta   d   \nonumber \\
{\bfXi} &&=  \xi - 2 \theta [{\mybar z}_2, {\mybar z}_1]_{\zeta^{*}}  \,.
\end{eqnarray}
Note that the ${\mybar z_a}$ are singlets under supersymmetry. 
The action of the deformed matrix model in  manifestly $\CQ=1$
supersymmetric form may be written as   
\begin{equation}
S_{\zeta} = \frac{1}{g^2} \Tr\int d\theta\,\Bigl[ -\half \bfl 
 \partial_\theta  \bfl -  \bfl \left[\mybar z_{a} ,\bfz_{a}\right] 
- \bfXi \ [ \bfz_{1}, \bfz_{2} ]_{\zeta}
\Bigr] \,.
\eqn{parent1}
\end{equation} 
The deformed matrix model  has a global symmetry group with a  subgroup 
$G_R= U(1) \times U(1)$.  We employ  $G_R$ to orbifold  the theory as in 
\cite{Cohen:2003xe}. We 
assign  the R-charges as in Table \ref{tab:tab1nc}. 

\setlength{\extrarowheight}{5pt}
\begin{table}[t]
\centerline{
\begin{tabular}
{|c||c|c|c|c||c|c|}
\hline
&$  \bfz_1 $&$\mybar  z_1 $&$ \bfz_2$&$\mybar  z_2$&$ \bfl$& $ \bfXi_{ab}$
\\ \hline
$ r_1 $&$ +1 $&$ -1 $&$ \,\ 0 $&$ \,\ 0 $&$ \,\ 0 $&$ -1 $
\\
$ r_2 $&$\,\ 0 $&$ \,\ 0 $&$ +1 $&$ -1 $&$ \,\ 0 $&$ -1 $
\\ \hline
 \end{tabular}
}
\caption{\sl The  $r_{1,2}$ charges  of the  fields of the $\CQ=1$  
deformed matrix theory
which  define the orbifold projection.\label{tab:tab1nc}}
\end{table}

Any of the fields of the deformed matrix model (with an   ${\bf r}$-charge 
vector given in Table \ref{tab:tab1nc}) satisfying  the orbifold constraint 
\Eq{orbifoldconstraint} may be written as in \Eq{neutralization} where 
the fields satisfying the neutral constraint are expressed as in 
\Eq{matrixlattice2}. For 
example, the superfield matrix ${\bfz_1}$ with $r$-charge vector $(1,0)$
 satisfying orbifold constraints \Eq{orbifoldconstraint} is written as 
\beq 
\bfz_1= {\widetilde \bfz_1}D_1 = \Bigl( 
\sum_{{\bf n} \in \mathbb{Z}_L^2} \Delta_{\bf n} \otimes 
\bfz_{1,\bf n} \Bigr) \, D_1
\eeq
and a similar expression can be written for the other matrices as well. The 
$k \times k $ matrix field $\bfz_{1,\bf n}$ is associated with the unit cell
 $\bfn \in \mathbb Z_L^2$. To understand where exactly a field  resides 
on the unit  cell, we need to examine its gauge transformation properties. 
Consider again for simplicity  ${\bfz_1}$. Under a gauge 
transformation,
 $\bfz_1 \rightarrow \widetilde g \bfz_1 \widetilde g^{\dagger} $  or 
equivalently 
\beq 
\widetilde \bfz_1 \rightarrow \widetilde g 
\widetilde \bfz_1 ( D_1
\widetilde g^{\dagger} D_1^{-1} )  
\eqn{gauget}
\eeq
where 
\beq
\widetilde g= \sum_{{\bf n} \in \mathbb{Z}_L^2} \Delta_{\bf n} \otimes 
g_{\bf n} 
\eeq
is the matrix symbol for gauge rotation. Using \Eq{difference} and 
\Eq{identities} and by taking the 
trace of both sides of \Eq{gauget}, we obtain 
\beq
 \bfz_{1,\bfn} \rightarrow g_{\bfn} \star  \bfz_{1,\bfn}
\star  g_{\bfn + {\hat {\bf e}_1}}^{\dagger}
\eeq
the star gauge transformation of $\bfz_{1,\bfn}$ where $g_{\bfn}$ is star 
unitary gauge rotation matrix.  Similarly, we have 
$ {\mybar z}_{1,\bfn} \rightarrow  g_{\bfn + {\hat {\bf e}_1}} 
\star  {\mybar z}_{1,\bfn} \star  
g_{\bfn}^{\dagger} $.  Both $\bfz_{1,\bfn}$ and ${\mybar z}_{1,\bfn}$ are 
residing on the link between sites $(\bfn, {\bfn + {\hat {\bf e}_1}})$ and 
transforming oppositely under star gauge transformation. Our choice of the 
ordering in \Eq{neutralization} was in fact  motivated by this convention. 
 The orientation of the link variables are determined by 
their ${\bf r}$-charges. To summarize, the star gauge transformation 
properties of the lattice fields are 
\begin{eqnarray}
&&\bfz_{a,\bfn} \rightarrow g_{\bfn} \star  \bfz_{a,\bfn}
\star  g_{\bfn + \ah}^{\dagger}  \nonumber \\
&&\mybar z_{a,\bfn} \rightarrow g_{\bfn + \ah} \star  \mybar z_{a,\bfn}
\star  g_{\bfn }^{\dagger}  \nonumber \nonumber \\
&&\bfl_{\bfn} \rightarrow g_{\bfn} \star  \Lambda_{\bfn}
\star  g_{\bfn }^{\dagger}  \nonumber \\
&&\bfXi_{\bfn} \rightarrow g_{\bfn + {\bf \hat e_1} +  {\bf \hat e_2} } 
\star  {\bf \Xi}_{\bfn}
\star  g_{\bfn }^{\dagger} \,.  
\eqn{gauget2}
\end{eqnarray}

From  
the supersymmetry transformation of the deformed matrix theory 
\Eq{parentsusy}, one can extract the supersymmetry transformation of the 
lattice theory by using the completeness and orthogonality \Eq{identities}
of the basis $\Delta_{\bf n}$ and the algebra 
\Eq{Dalgebra} of translation $D_a$ matrices.   
In terms of individual components, the supersymmetry transformations of 
the lattice fields   can be found as 
\begin{equation}
\begin{aligned}
\delta z_{a,\bfn} &= i\sqrt{2}\,\eta\psi_{a,\bfn} \\
\delta \mybar z_{a,\bfn} &= 0\\
\delta\psi_{a,\bfn}&=0 \\
\delta\xi_{\bfn} &=-2i \eta (\mybar
  z_{2,\bfn+ {\bf \hat e_1} } \star \mybar z_{1,\bfn} - 
\,\zeta^{*} \, \omega_{m_1m_2q}^{-1} \,\,
  \mybar z_{1,\bfn+ {\bf \hat e_2} } \star \mybar z_{2,\bfn})\\
\delta\lambda_\bfn &=  \eta d_{\bfn}  \\ 
\delta d_\bfn &= 0 \,,
\eqn{dfieldsbnc}\end{aligned}
\end{equation} 
 where the extra phase factor $\omega_{m_1m_2q} $ arises because 
of noncommutativity of translations in orthogonal directions. 
 Note that $\mybar z_{a, \bfn}$  are 
supersymmetric singlets. These transformation laws lead 
to  supersymmetric multiplets
\begin{equation}
\begin{aligned}
{\bf \Lambda}_{\bfn}&= \lambda_{\bfn} -i\theta  d_{\bfn} \ ,\\
 &&\\
{\bfz}_{a,\bfn} &= z_{a,\bfn} + \sqrt{2}\,\theta \,\psi_{a,\bfn}\ ,\\
 &&\\
{\bf \Xi}_{\bfn}&= \xi_{\bfn} -  2\theta\,\, (\mybar
  z_{2,\bfn+ {\bf \hat e_1} } \star \mybar z_{1,\bfn} - 
\,\zeta^{*} \, \omega_{m_1m_2q}^{-1} \,\,
  \mybar z_{1,\bfn+ {\bf \hat e_2} } \star \mybar z_{2,\bfn})
 \ .
\end{aligned}
\end{equation}

Setting $\zeta \, \omega_{m_1m_2q}=1$, we express 
the lattice action for the noncommutative  $\CN=(2,2)$ theory 
in manifestly $\CQ=1$
supersymmetric form as 
\begin{equation}
\begin{aligned}
S = \frac{1}{g_{nc}^2} \sum_{\bfn \in \mathbb Z_L^2} \tr\int d\theta\,\Bigl[& -\half \bfl_\bfn 
\star \partial_\theta  \bfl_\bfn -  \bfl_\bfn \star \left(\mybar
    z_{a,\bfn-\ah} \star 
    \bfz_{a,\bfn-\ah} - \bfz_{a,\bfn} \star \mybar
    z_{a,\bfn}\right) \\
&
-  \bfXi_{\bfn} \star ( \bfz_{1,\bfn} \star \bfz_{2,\bfn
  +{\bf \hat e_1}} - \bfz_{2,\bfn} \star \bfz_{1,\bfn + {\bf \hat e_2} })
\Bigr]
\eqn{ssact3nc} \,,
\end{aligned}
\end{equation} 
where  $g^2_{nc}= g^2 \frac{m_1}{m_2} $ is the dimensionful coupling 
\footnote{In our normalization, the mass dimension for fields, coupling 
constant and Grassmann coordinate are:
$$[z]=[\mybar z]=[\rma^{-1}]=1, \; [d]=2, \;\;  [{\rm fermions}]=3/2, \;\;
[g^2_{nc}]=4,\;\; [\partial_\theta]= - [\theta]= 1/2.$$} 
Substituting the superfields and integrating over the Grassmann coordinate, 
we can  obtain the action in components.  

The action has a classical 
moduli space (flat directions in the field configuration 
along which  the bosonic part of the action  vanishes) 
 for $\zeta=\omega_{m_1m_2q}^{-1}$ analogous to its 
commutative counterpart \cite{Cohen:2003xe}. For any other value of $\zeta$,
 there is generically no moduli space along which both moduli fields are 
nonvanishing, and a sensible continuum limit does not exist. 
We will examine these cases in subsection \ref{subsec:moral}. 
The flatness condition can 
be written as  
\begin{eqnarray}
&& \d_{\bfn} = \mybar    z_{a,\bfn-\ah} \star 
       z_{a,\bfn-\ah} -    z_{a,\bfn} \star \mybar
    z_{a,\bfn}  = 0 \nonumber \\
&&  z_{1,\bfn} \star z_{2,\bfn
  +{\bf \hat e_1}} - z_{2,\bfn} \star z_{1,\bfn + {\bf \hat e_2} }=0
\eqn{moduli}
\end{eqnarray} 
which are analogs  of the D-flatness and  F-flatness conditions that 
show up in supersymmetric gauge theories in higher dimensions. 

As in the 
commutative counterpart, we consider a particular configuration 
in moduli space respecting the $U(k)$  global symmetry, namely
 \beq 
z_{a,\bfn}=\mybar z_{a, \bfn}= \frac{1}{\rma\sqrt{2}} {\bf 1}_k    \qquad a=1,2
\eqn{point}
\eeq
where $\rma$ is a parameter with dimension of length. 
We expand the action about 
the point \Eq{point} by identifying the lattice spacing with $\rma$.   

\subsection{Continuum Limits}
In any noncommutative field theory the quadratic part of the action
must be exactly the same as the quadratic part of its commutative 
counterpart,
 both for the lattice and  the continuum, which amounts to saying 
that  the free propagators are the  same. 
The difference starts with the momentum dependent phase factors 
attached to interaction vertices. Hence, our analysis of the tree level 
is quite similar to \cite{Cohen:2003xe} and in fact, the identification of 
continuum fields in terms of lattice fields is exactly the same.  
 Here, our emphasis will be on different 
characteristics of the continuum limit of noncommutative theories with respect 
to their commutative counterparts. 

To construct the possible continuum limits of the non-commutative 
SYM  theories, 
we first  obtain the continuum Moyal $\star$ product  from the lattice 
$\star$ 
product \Eq{latticeMoyal} and identify the dimensionful  noncommutativity 
parameter.    
Let $
{\bf w} = {\bf s} \rma,\,    {\bf x} = {\bf m} \rma, \,    
{\bf y} = {\bf n} \rma $ be dimensionful length. We need to replace  the 
summation in   \Eq{latticeMoyal} with integration.  But naively replacing 
the two  summations with two  integrations would simply be wrong since the 
summations are not over all lattice points. We can take without loss of 
generality one of the summations to be over all lattice points. Then 
we substitute
$
\sum_{\bf m} {\rma}^2 \rightarrow \int d^2x \,.  
$
In the second summation, ${\bf n}$ runs over 
the set $\frac {L\theta'}{2} \mathbb Z \times 
\frac {L\theta'}{2} \mathbb Z $, a sublattice 
of integers,  modulo periodicity of the lattice.  Hence,  the sum 
with a prime has to be replaced with  
\beq
{\sum_{{\bf n} \in  \mathbb Z^2_L}}^{\prime}
{\rma}^2 \rightarrow  (\frac {2}{L \theta' })^2 \int d^2y  
\eeq 
With these deliberations in mind, we do obtain the continuum Moyal 
star product from the lattice star product 
\Eq{latticeMoyal} as
\begin{eqnarray}
\Phi_1({\bf w}) \star  \Phi_2({\bf w}) =&&  \frac {1}{L^2\rma^4 } 
\frac{4}{L^2 \theta'^2} 
\int \int  \, d^2x \,  d^2y  \,\, \Phi_1({\bf x}) \,  \Phi_2({\bf y}) 
\,\,  e^{-2i \frac{2 \pi}
{L^2 \theta' {\rma}^2} ({\bf x} - {\bf w})\wedge  ({\bf y} - {\bf w})}  \\ 
=&&
\int \int  \, d^2x \,  d^2y  \,\, \Phi_1({\bf x}) \,  \Phi_2({\bf y}) 
\,\, K({\bf x} - {\bf w}, {\bf y} - {\bf w})   
\end{eqnarray}
where the kernel of the integral is defined as 
\beq
K({\bf x} - {\bf w}, {\bf y} - {\bf w})= \frac {1}{\pi^2 {\rm det}\Theta'}
 e^{-2i  ({\bf x} - {\bf w})_i (\Theta'^{-1})_{ij} ({\bf y} - {\bf w})_j}
\eeq 
We identify the dimensionful  noncommutativity parameter 
$\Theta_{ij}^{\prime}$ as
\beq
\Theta'_{ij} = \frac {L^2 {\rma}^2  \theta' }{2 \pi} \epsilon_{ij} \,,
\eqn{Theta'}
\eeq
where  $\epsilon_{12}= - \epsilon_{21}=1$. 

\noindent{\it Finite volume }

There are two  types of continuum limit that one can consider with a finite 
volume and 
finite two dimensional coupling $g_2^2 = g_{nc}^2 \rma^2$. 
One choice is  finite noncommutativity and   the other is zero  
noncommutativity. From the relation \Eq{Theta'}, we observe that 
infinite noncommutativity is not compatible with the finite 
volume limit because $\theta' \in [0,1)$ .  
Let $\alpha$ be a parametrically small number.   Consider the scaling      
\beq
\rma \sim   \alpha, \qquad  L \sim \frac{1}{ \alpha}
\,\,,   \qquad 
g^2_{nc}  \sim  \frac{1}{\alpha^2}, \qquad 
\eqn{scaling1}
\eeq
which yields finite volume and finite two dimensional coupling. 
 But there are two distinct outcomes  for dimensionful 
noncommutativity parameter $\Theta'$ depending on the scaling of the 
dimensionless parameter $\theta'$:
\beq
{\rm Vol}= (\rma L)^2 \sim 1, \,\,\,\,\, 
g_2^2   \sim  1, \,\,\,\,\,
\Theta^{\prime} \sim  
\left\{ \begin{array}{cc}
      \alpha &  \, {\rm for} \,\, \theta' \sim \alpha  \\
      1       & \, {\rm for} \,\, \theta' \sim 1 
\end{array} \right.  \,\,.
\eqn{dimen1}
 \eeq
The $\theta'$ on the lattice is defined modulo ${\mathbb Z}$, it is a rational 
number between zero and one with a minimum nontrivial value 
$\theta'=\frac{2}{L}$ and a generic value  as in \Eq{theta'}. If $\theta'$ 
is kept fixed as we take the limit of large $L$ as in \Eq{scaling1}, we 
obtain a finite noncommutativity parameter $\Theta'$ in the continuum. If 
$\theta'$ vanishes in the limit \Eq{scaling1}, then we obtain  zero 
noncommutativity parameter in the continuum with fixed volume.       

\noindent{\it Infinite volume}

The continuum limit in which the volume is taken  to infinity is obtained 
from the scaling 
\beq
\rma \sim   \alpha, \qquad L \rightarrow \frac{1}{\alpha^2}
,   \qquad 
g^2_{nc}  \rightarrow  \frac{1}{\alpha^2} \,\, . 
\eqn{scaling2}
\eeq
Hence the continuum parameters and volume scale as 
\beq
{\rm Vol} 
\sim \frac{1}{\alpha^{2}}, \,\,\,\,\,\qquad  
g_2^2 
\sim  1, \,\,\,\,\, \qquad
\Theta' \sim 
  \left\{ \begin{array}{cc}
      1 &  \, {\rm for} \,\, \theta' \sim \alpha^2  \\
      \frac{1}{\alpha^2}       & \, {\rm for} \,\, \theta' \sim 1 
\end{array} \right. 
\eqn{dimen2} \,\,.
 \eeq
Notice that the limit $\theta' \rightarrow  0$  which leads to zero 
noncommutativity parameter in  finite volume leads 
to a finite dimensionful noncommutativity in the infinite volume. 
Similarly, 
the scaling  $\theta' \sim 1 $ which leads to the finite noncommutativity 
on finite volume 
leads to infinite dimensionful noncommutativity in the infinite 
volume because of the relation \Eq{Theta'}. 

\noindent{\it Continuum: tree level} 

By expanding the action \Eq{ssact3nc} 
about 
the point \Eq{point} in moduli space, we  obtain the action of the  
continuum noncommutative $\CN=(2,2)$ SYM theory. Splitting 
the fluctuations of the link field into its hermitian and antihermitian 
parts, we have 
\beq
z_{a}= \frac{1}{\sqrt 2 \rma}{\bf 1}_k + \frac{s_a + i v_a}{\sqrt 2} 
\eeq
where $s_a$ and $v_a$ are hermitian $k\times k $ matrices.  The fields 
$s_a$ are identified as the scalar in the continuum, and   $v_a$ are the 
gauge fields. The fermions residing on the links and sites combine to form 
the Dirac spinors of the two dimensional theory. These results follows 
easily by using the property  \Eq{const}  of the star product. Essentially, 
the analysis of the quadratic part of the action in the noncommutative theory 
reduces to the one in ordinary theory. 
 The spectrum of the fermions is the same as
the spectrum of scalars owing to exact supersymmetry on the lattice, and there 
are no doublers in the formulation \cite{Cohen:2003xe}.    

The continuum star gauge transformations follow from the smooth gauge 
transformation on the lattice. Consider for example the transformation 
of $\bfz_{a,\bfn}$ on the lattice \Eq{gauget2}. By expanding the gauge 
transformation at $\bfn+ \ah$ as 
$g({\bf x} + \ah \rma)=  g({\bf x}) + \rma \partial_a g({\bf x})+ O(\rma^2)$ 
and 
using star unitarity $g_{\bfn}  \star g_{\bf n}^{\dagger}=1$, we 
obtain  the desired gauge transformation  in the continuum  
\begin{eqnarray}
&& s_a({\bf x})  \rightarrow g({\bf x})   \star s_a({\bf x})
\star  g^{\dagger}({\bf x})  + O(\rma), \nonumber \\
&&  v_a({\bf x})  \rightarrow g({\bf x}) \star  v_{a}({\bf x})
\star  g^{\dagger}({\bf x}) -i  g({\bf x})\star \partial_a 
g^{\dagger}({\bf x}) + O(\rma)
\end{eqnarray}
where the scalars transform homogeneously and the vector potential 
transforms inhomogeneously. 
Similarly, the fermionic matter transforms in the  adjoint representation 
of the 
group $U(k)$ up to $O(\rma)$ corrections. 

The action of the continuum $\CN=(2,2)$ SYM target theory with gauge group 
$U(k)$ may be written as  
\begin{equation} 
  S =  \frac{1}{g_2^2}  \,\int d^2x \;  \tr
  \Biggl( D_m s_a \star D_m s_a 
+ \half v_{12}\star v_{12} 
- \half
 ( s_1 \star  s_2\, -s_2 \star s_1 )^2 + \rm{fermions} \Biggr) 
  \eqn{targ2nc}
\end{equation} 
where  $D_ms_a = \partial_m s_a  + i
(v_m \star s_a - s_a \star v_m )$ is the star covariant derivative of scalar 
$s_a$, and $v_{12}= \partial_1 v_2 - \partial_2 v_1 + i (v_1 \star v_2 - 
v_2 \star v_1)$ 
is the field strength.
An important distinction between the noncommutative and  ordinary SYM 
theory  is the fact that 
the $U(1)$ sector  in the noncommutative theory is an interacting field 
theory and does not decouple from the $SU(k)$ sector \cite{Armoni:2000xr}. 
 There are both  three photon, three gluon as well as one  photon, 
two gluon vertices in a perturbative expansion. We will discuss the continuum 
of the quantum theory after discussing the $SL(2,\mathbb Z)$ duality of a 
noncommutative 
gauge theory to  a commutative gauge theory with flux. 

\subsection{The moral  for the deformation} 
\label{subsec:moral}
In order to reach the lattice action for noncommutative field 
theory, we started with a deformed matrix action with an a priori 
undetermined parameter $\zeta$. Then we set 
$\zeta\omega_{m_1m_2q}=1$ to obtain the lattice action for the 
noncommutative theory. Here, we will argue that 
$\zeta=\omega_{m_1m_2q}^{-1}$  
is the only choice which yields a useful moduli space in the sense 
that  it  supports the 
lattice  action for noncommutative $\CN=(2,2)$ SYM and possesses a  sensible 
continuum.

For an arbitrary value of the  deformation parameter $\zeta$, the second 
equation determining moduli space in \Eq{moduli} 
 would  take the form 
\begin{eqnarray}
 && z_{1,\bfn} \star z_{2,\bfn +{\bf \hat e_1}} -  \zeta \omega_{m_1m_2q} \,\,
 z_{2,\bfn} \star z_{1,\bfn + {\bf \hat e_2}  }=0\,, 
\eqn{modul2}
\end{eqnarray} 
where  $z_{a, \bfn}$ are $k\times k$ matrices and $\omega_{m_1m_2q}$ 
is $(m_1m_2q)$-th  root of unity. There is a priori no relation between 
the integers $ {m_1m_2q}$ and $k$. 

For  generic values of $k$ and $(m_1m_2q)$, consider the 
$\bfn$-independent 
configurations of moduli fields satisfying  the  
matrix equation \Eq{modul2}.  By using \Eq{const}, we can rewrite the 
moduli equation as  
\begin{eqnarray}
 && z_{1}  z_{2} -  \zeta \omega_{m_1m_2q} \,\,
 z_{2}  z_{1}=0 \,, 
\eqn{moduli2}
\end{eqnarray} 
where $z_a$ is a rank $k$, ${\bfn}$-independent  matrix. 
Taking trace of both sides, we obtain 
\beq
(1- \zeta \omega_{m_1m_2q}) \tr(z_1z_2)=0\,. 
\eeq
If  $\tr(z_1z_2) \neq 0$, the only way to have a solution is the vanishing 
of the first term,  requiring $\zeta \omega_{m_1m_2q} = 1$. 

For generic values of the parameter $\zeta \omega_{m_1m_2q}$,  
there is no solution at all to matrix equations \Eq{moduli2} unless  
one of the $z_{a}= 0$. That means one of the moduli 
fields 
is frozen at the origin of the moduli space and can not move. 
But in order to take the 
continuum limit, we have to take both $\langle z_{1,\bfn}\rangle$ and $
\langle z_{2, \bfn} \rangle$  
to large values of moduli (which corresponds to small lattice spacing)  
proportional to the identity.  This can not be achieved 
under the given circumstances. 

For some  special values of parameters, say $\zeta=1$ and  $k=(m_1m_2q)$, 
there is a moduli 
space where both $z_{a}$ are nonvanishing and satisfying  $\tr(z_1z_2) = 0$.
 But it is a noncommutative  moduli space and 
the identity matrix is not part of it.  
The noncommutative  moduli space can be realized  with the matrices 
\beq 
z_{1}=\frac{1}{\rma} U_k, \qquad   z_{2}=\frac{1}{\rma} V_k 
\eeq
where $U_k \,  {\rm and} \,  V_k$  are  defined in \Eq{UV} and 
$\rma$ is a parameter with inverse mass dimension. 
Expanding the lattice action  
around such a background configuration does not 
lead to a  sensible  continuum action.  
For example, it  contains   terms such as  
\beq
 \sum_{\bfn}  \tr |\frac{1}{\rma}( U_k \star z_{2, \bfn +{\bf \hat e_1} }  - 
\omega_k  z_{2, \bfn} \star U_k ) |^2 .
\eeq
Even for smooth configurations of the fields $z_{2,\bfn}$, this term  
(as well as  other terms in the action)  tends to infinity as we take 
$\rma \rightarrow 0$ limit.  Thus, even though there is a 
noncommutative  moduli space,  there is no sensible 
continuum limit that can emerge anywhere in this moduli space. Moreover, 
even at finite lattice spacing, we know that the free spectrum of a 
noncommutative  theory has to be identical to its commutative counterpart. 
This is not realized within  the noncommutative moduli space.

With the deformation  we performed to  the  parent  matrix theory, 
we do obtain 
a lattice action for noncommutative SYM theory which possesses  a sensible 
continuum limit.  
In fact,  the free spectrum  of the  noncommutative 
theory turns out to be identical to its ordinary counterpart. This 
is expected since the difference between the two theories at the 
perturbative level starts with momentum 
dependent phase associated to the cubic and higher order interaction vertices. 
Thus,  the 
perturbative spectrum of the $\CN=(2,2)$ noncommutative SYM coincides 
with the ordinary SYM examined in \cite{Cohen:2003xe}. 

\section{Nonperturbative T-duality (Morita Equivalence)}
\label{sec:Tduality}
In section \ref{sec:reg}, we constructed a nonperturbative regularization 
for   noncommutative $\CN=(2,2)$  
supersymmetric gauge theories on torus. In this section, we will 
nonperturbatively prove that 
nontrivial topological sectors (with an integer 
 background nonabelian magnetic flux)  of a $U(k)$ $\CN=(2,2)$  
SYM theory on a 
commutative torus
are equivalent  to  a noncommutative SYM theory on a larger size torus  
with a reduced gauge group. The noncommutativity parameter is uniquely 
determined by the integer magnetic flux. Our proof also implies that 
noncommutative SYM  theories can be regularized by means of an ordinary 
SYM theory with 't Hooft flux.  
The discussion in this section is not  specific to $\CN=(2,2)$, 
it also applies to   $\CN=(4,4)$ and   $\CN=(8,8)$ theories.

Let us start with the lattice action for a $U(k)$ $\CN=(2,2)$ supersymmetric 
gauge theory on an $L \times L $ lattice.  
The action for this theory may be written as  in \cite{Cohen:2003xe}  
\begin{equation}
\begin{aligned}
S = \frac{1}{g^2} \sum_{\bfn \in \mathbb Z_L^2 }\tr_{(k)}\int d\theta\,\Bigl[& -\half \bfl_\bfn 
 \partial_\theta  \bfl_\bfn -  \bfl_\bfn  \left(\mybar
    z_{a,\bfn-\ah} 
    \bfz_{a,\bfn-\ah} - \bfz_{a,\bfn}  \mybar
    z_{a,\bfn}\right) \\
&
-  \bfXi_{\bfn}  ( \bfz_{1,\bfn}  \bfz_{2,\bfn
  +{\bf \hat e_1} } - \bfz_{2,\bfn}  \bfz_{1,\bfn +{\bf \hat e_2} })
\Bigr]
\eqn{commutative}
\end{aligned}
\end{equation}
or equivalently as a matrix theory action 
\begin{equation}
S = \frac{1}{g^2} \Tr_{(L^2)} \tr_{(k)}\int d\theta\,\Bigl[
-\half \widetilde 
\bfl 
 \partial_\theta \widetilde  \bfl - \widetilde \bfl  [D_a^{-1}
\; \widetilde{\mybar z}_{a},\; 
   \widetilde \bfz_{a}D_a]
- \widetilde \bfXi [ \widetilde \bfz_{1} ( D_1  
\widetilde \bfz_{2} D_1^{-1}) -  
  \bfz_{2} (D_2 \widetilde \bfz_{1} D_2^{-1}) ]
\Bigr]
\eqn{commutativematrix}
\end{equation} 
where we have used the commuting rank-$L^2$ basis matrices to express 
the lattice as a large matrix, 
and $D_a$ are the commuting finite difference operators. The matrices with 
tilde are rank-$L^2k$ and are  defined as in \Eq{matrixlattice2}.    
The  $\Tr_{(L^2)}$
is the trace over the lattice basis and  $\tr_{(k)}$ is over the $U(k)$ gauge 
group. The subscripts of trace denote the rank of the corresponding matrices. 
 
To construct the different topological sectors of the theory,  
we impose on the superfields   the following twisted boundary 
conditions \cite{'tHooft:1977hy, 'tHooft:1979uj}
\begin{eqnarray}
&&
{\bf \Lambda}_{\bfn + L \ah}= \Gamma_{a,\bfn} \; {\bf \Lambda}_{\bfn} 
\; \Gamma_{a,\bfn}^{\dagger} \nonumber \\ 
&&
\bfz_{a, \bfn + L \bh}= \Gamma_{b,\bfn} \; \bfz_{a,\bfn} 
\; \Gamma_{b,\bfn + \ah  }^{\dagger} \nonumber \\
&&
\mybar z_{a, \bfn + L \bh}= \Gamma_{b,\bfn + \ah} \;
\mybar z_{a,\bfn} \; \Gamma_{b,\bfn }^{\dagger} \nonumber \\
&&
{\bf \Xi}_{\bfn + L \ah}= \Gamma_{a,\bfn+ {\bf \hat e_1} + {\bf \hat e_2}}
 \; {\bf \Xi}_{\bfn} 
\; \Gamma_{a,\bfn}^{\dagger}  
\eqn{twisted}
\end{eqnarray}
where $L$ is the periodicity of the lattice in two orthogonal directions.  
The twisted boundary conditions state that after making a cycle on the 
discrete 
torus 
(in either direction)
the field matrix at site $\bfn$ does not necessarily turns back to 
its original value but to a configuration which is related to it  
by a gauge transformation. Hence, the fields are not single valued and 
should be considered as living on the covering space ${\mathbb Z}^2$ 
of the discrete torus ${\mathbb Z_L}^2$. But the covering space 
${\mathbb Z}^2$ 
is in fact bigger then what we need, as  will be shown below. 
The consistency of the field configurations at ${\bfn + L \ah+ L \bh}$ 
requires
the gauge rotations  $\Gamma_{a, \bfn}$ 
to satisfy the algebra 
\beq
\Gamma_{a, \bfn + L \bh} \, \Gamma_{b, \bfn} = e^{2 \pi i Q_{ab}/k}  
\Gamma_{b, \bfn + L \ah} \,
\Gamma_{a, \bfn}
 \eeq
where $Q_{ab} \in \mathbb Z $ is the 't Hooft flux and 
$e^{2 \pi i Q_{ab}/k}$ is in the center ${\mathbb Z}_k$ of $SU(k)$ group.
 In 
particular, we can choose the gauge rotations to be spacetime independent, 
with $\Gamma_{a, \bfn}= \Gamma_{a}$. In this case,   
 the above algebra turns into 
\beq
\Gamma_{a} \, \Gamma_{b} = 
e^{2 \pi i Q_{ab}/k}  \Gamma_{b} \,\Gamma_{a}
\eqn{Gamma} \,\, .  
\eeq
't Hooft also proved that there are $k$ distinct, non-gauge equivalent 
 choices of boundary conditions, one for each choice of 
$Q_{ab} \in \mathbb Z $ modulo $k$.  The flux matrix $Q_{ab}$ is an 
antisymmetric matrix and can be written as  $Q_{ab}= \epsilon_{ab}\,q$ where 
$\epsilon_{12}=-\epsilon_{21}=1$. The  $e^{2 \pi i q/k}$ factor  
either generates the center ${\mathbb Z}_k$ or a proper subgroup of it. 
Let the  
greatest common divisor of $q$ and $k$ be $gcd(q,k)=k_0$. Then we can write 
$q=k_0q_1$ and  $k=k_0k_1$ where $q_1$ and $k_1$ are coprime. Thus, we can 
express  the generator   
\beq
e^{2 \pi i \frac{q}{k}}= e^{ 2 \pi i \frac{q_1}{k_1}} \qquad {\rm with} \qquad
gcd(q_1,k_1)=1, 
\eeq
which generates a   ${\mathbb Z}_{k_1}$ subgroup of  ${\mathbb Z}_k$. 
A convenient representation for the  algebra \Eq{Gamma} is   (in terms of 
the rank-k matrices) \cite{vanBaal:1985na, Lebedev:1985gp} 
\begin{eqnarray}
&&\Gamma_1=U_{k_1}^{\dagger} \otimes {\bf 1}_{k_0}  \nonumber \\ 
&&\Gamma_2=V_{k_1}^{q_1} \otimes {\bf 1}_{k_0} \ .
\end{eqnarray}

Now, let us turn back to the covering space. It is easy to see that after 
making  $k_1$ cycles in either direction on the discrete torus, the field 
configuration turns back to its original value. Thus, it is sufficient to 
consider  $k_1^2$-fold cover ${\mathbb Z}_{Lk_1}^2$  
of the discrete torus  ${\mathbb Z}_{L}^2$. The fields are periodic 
on the    $Lk_1 \times Lk_1$ lattice. The covering space will be essential 
for constructing the duality. 

Next, we  will show that  the commutative $U(k)$ supersymmetric lattice gauge 
theory \Eq{commutative} on  ${\mathbb Z_L^2}$ lattice 
 with twisted boundary conditions 
 \Eq{twisted} is equivalent to a 
noncommutative SYM with reduced rank $k_0$ on the covering space  
${\mathbb Z_{Lk_1}^2}$ with periodic 
boundary conditions. The action \Eq{commutative} includes a sum over all 
lattice points ${\bfn \in \mathbb Z_L^2}$  with twisted boundary condition 
 \Eq{twisted}. 
As a first step, we can regard the sum over all points on the covering space 
${\bfn \in \mathbb Z_{Lk_1}^2}$ and declare the boundary conditions 
\Eq{twisted} 
to be  a property that the fields living on ${ \mathbb Z_{Lk_1}^2}$ must 
satisfy.  This simple observation will lead  us to rewrite 't Hooft's twisted 
boundary conditions as an orbifold with discrete torsion.   

We can express the lattice action \Eq{commutative} (or
 \Eq{commutativematrix}) with twisted boundary conditions as 
a matrix theory action on covering space as 
\begin{equation}
S = \frac{1}{g^2} \Tr_{(Lk_1)^2} \tr_{(k)} \int d\theta\,  
\Bigl[ -\half \widetilde \bfl 
 \partial_\theta  \widetilde \bfl -  \widetilde \bfl \left[D_a^{-1}   
 \widetilde {\mybar z}_{a} ,\widetilde \bfz_{a} D_a \right] 
- D_2^{-1}D_1^{-1} \widetilde \bfXi [ 
\widetilde \bfz_{1} D_1  , \widetilde \bfz_{2} D_2 ]
\Bigr]
\eqn{matrixaction}
\end{equation} 
where the matrices in the action are  rank-$(Lk_1)^2 k$ matrices and 
the $D_a$  are the commuting 
 lattice shift operators on covering space 
acting as in \Eq{difference}.

In order to write down the  twisted boundary conditions in terms of 
matrix symbols on the covering space, we define the matrix symbol for 
a gauge rotation matrix as 
\beq
\widetilde \Gamma_a =  \sum_{\bfn \in \mathbb Z_{Lk_1}^2} 
\Delta_{\bfn} \otimes \Gamma_a = {\bf 1}_{(Lk_1)^2} \otimes \Gamma_a \,.
\eeq
Thus, the boundary conditions \Eq{twisted}
can be captured in terms of a generic  matrix  
symbol $\widetilde \Phi$ as  
\beq
(D_a)^{L} \widetilde \Phi (D_a^{\dagger})^{L}
= \widetilde \Gamma_a \widetilde \Phi \widetilde \Gamma_a^{\dagger} 
\eqn{matrixconstraint}\,.
\eeq
For example, consider $\bfz_1$ which has an r-charge vector $(1,0)$.  Using 
the recipe \Eq{neutralization}, we write it as  $\bfz_1 = \widetilde \bfz_1
D_1$ where $\widetilde \bfz_1$ satisfies \Eq{matrixconstraint}. Thus,  $\bfz_1$
obeys  
\begin{eqnarray} 
(D_a)^{L}  \bfz_1  (D_a^{\dagger})^{L}
&=& \widetilde \Gamma_a    \bfz_1 (D_1^{-1} \widetilde \Gamma_a^{\dagger} D_1) 
\end{eqnarray}
By using \Eq{difference} and \Eq{identities}, and taking trace 
$\Tr_{(Lk_1)^2}$ 
of both sides, we do recover the  twisted boundary conditions given in 
\Eq{twisted}.  Instead, 
we  can rewrite the constraints as
 \beq
 \widetilde \Phi 
= (D_a^{\dagger})^{L}  \widetilde \Gamma_a \,\widetilde \Phi \, 
\widetilde \Gamma_a^{\dagger}
 (D_a)^{L} = \Omega_a^{\dagger} \, \widetilde \Phi \, \Omega_a
\qquad  {\rm with} \qquad \Omega_a = \widetilde \Gamma_a^{\dagger}(D_a)^{L}
 \eqn{matrixconstraint2} \,,
\eeq
which is the neutral orbifold constraint given in \Eq{neutralconstraint}, 
and from which we can obtain \Eq{orbifoldconstraint}, the starting point 
of our construction.
The orbifold matrices form an  $(Lk_1)^2k$ dimensional projective 
representation of the cyclic group $ \mathbb Z_{k_1}$. 
Thus the orbifold  is a 
$  \mathbb Z_{k_1} \times   \mathbb Z_{k_1}$ orbifold where 
orbifold matrices satisfy
\beq 
\Omega_1 \Omega_2 = e^{2\pi i \theta} \Omega_2 \Omega_1 \qquad {\rm with}
\qquad \theta=  \frac{q_1}{k_1} \,.
 \eeq
The noncommutativity of orbifold matrices is purely because of their 
't Hooft $\Gamma_a$ gauge rotation component. An 
explicit formula for $\Omega_a$'s can be given as  
\beq
 \Omega_1 = V_{L'}^L \otimes {\bf 1}_{L'} \otimes U_{k_1}
\otimes {\bf 1}_{k_0} \,\, ,
\qquad 
 \Omega_2= {\bf 1}_{L'} \otimes  V_{L'}^L \otimes (V_{k_1}^{\dagger})^{q_1}
\otimes {\bf 1}_{k_0} \,.
\eeq
where $L'=Lk_1$.
We have expressed 
 't Hooft's twisted boundary conditions as
an orbifold constraint with discrete torsion  on a 
parent matrix theory of rank $(L')^2k$, and we can use techniques from 
section \ref{sec:nclattice}  to solve it. This interpretation also makes the 
physical meaning of the discrete torsion clear. Under the given circumstances, 
the discrete torsion is the ratio of background magnetic flux  to the rank 
of the gauge group.  Equivalently, it is the ratio of the topological charges 
of the theory:
\beq
\theta = \frac {q}{k}= \frac{q_1}{k_1}= \frac{{\rm ch}_1}{{\rm ch}_0} 
\eeq
where ${\rm ch}_0$  is the zeroth Chern character (which  counts the rank of 
the 
gauge group) and ${\rm ch}_1$  is  first Chern character (which 
determines the 
net integer background flux). Hence the noncommutativity parameter is
a topological quantity. 

The set of constraints \Eq{matrixconstraint} had  been solved rigorously  
in 
\cite{Ambjorn:1999ts} in a similar context for pure Yang-Mills theory. 
The main idea  is to look for a solution in the $k_1^2$-fold 
 covering space $\mathbb Z_{Lk_1}^2$,  $k_1$ 
replicas of the lattice in  each direction. 
As we realized,  this in turn requires being tolerant to the rank 
of the deformed parent  matrix model.  
The dual  noncommutative SYM gauge theory does not originate from the matrix 
theory that we used to obtain  SYM with magnetic flux $Q$. The rank of the 
parent matrix theory which give rise to the ordinary SYM theory with flux is  
$L^2k$, whereas the rank of the parent  which generates the dual 
noncommutative theory  is $(Lk_1)^2k$. The difference of the ranks of the 
parent matrix theories comes from the 
fact that the dual theory 
lives on the covering space. However, this is 
not a   problem.
  As we will show later, the total number of elementary degrees of freedom 
in the commutative theory with flux is equal to the number in the 
noncommutative 
theory by a simple counting of momentum modes in both theories. We use the 
results of section \ref{sec:nclattice} to find the solution to the constraints 
\Eq{matrixconstraint} and to obtain the action of the dual theory.

The full set of 
 commutants of the orbifold matrices $\Omega_a$ are the basis matrices 
$Z_a^{\prime} \otimes {\bf 1}_{k_0}$ of the noncommutative lattice  
and are given as  
\beq
Z_1^{\prime}= U_{L'} \otimes {\bf 1}_{L'} \otimes (V_{k_1}^{q_1})^b \,,\qquad  
Z_2^{\prime}= {\bf 1}_{L'}  \otimes  U_{L'} \otimes (U^{\dagger}_{k_1})^b  
\eeq 
where $b$ is an integer. The condition for the  basis matrices to commute 
with orbifold matrices requires  $\frac{1-q_1b}{k_1}=a$ to be an integer. 
The equation 
\beq
1 =k_1 a  + q_1 b    
\eeq
is guaranteed to have a solution  for some 
integers $(a,b)$  by construction since 
 $k_1$ and $q_1$ are coprime as in \Eq{Diophantine}.  

The complete set of solutions to the orbifold conditions is given in terms  
of momentum basis matrices on  the dual torus as 
\beq
\{(J_{\bf p}^{\dagger})^{\prime}= e^{\pi i \theta' p_1 p_2} 
(Z_2^{\prime})^{p_2}
(Z_1^{\prime})^{p_1} \,|\, {\bf p} \in {\mathbb Z}_{L'}^2 \}
\eeq
with  $(Z_a^{\prime})$ obeying the commutation relations 
\beq 
(Z_1^{\prime}) (Z_2^{\prime}) = e^{-2 \pi i \theta'} (Z_2^{\prime}) 
(Z_1^{\prime}) \, , \qquad 
\theta' = \frac{q_1b^2}{k_1} = \frac{b}{k_1} \,\, {\rm modulo} \,\, {\mathbb Z}
\eeq
with noncommutativity parameter $\theta'$.  The noncommutativity 
of the new basis is due to the existence of nonvanishing 't Hooft flux. 
 The parameter $\theta'$ is uniquely 
determined  by $\frac{b}{k_1}$ where the integer $b$ is the multiplicative 
inverse 
of the magnetic flux $q_1$ modulo $k_1$, that is  $bq_1=1$ modulo $k_1$. 
In fact, the solution to the Diophantine equation is  unique up to  
periodicities. Let $(a,b)$ be a 
solution, then so is $(a+mq_1, b-mk_1)$  where $m$ is any integer.
Thus,
\beq
  (a,b) \cong (a + mq_1, b-mk_1)
\eeq  
defines these congruences.

The most general solution to \Eq{matrixconstraint} may be written in 
the new momentum  basis as  in \Eq{matrixlattice} 
\beq
\widetilde \Phi = \sum_{{\bf p} \in \mathbb Z_{L'}^2 }  
(J_{\bf p}^{\dagger})^{'}
\otimes \Phi_{\bf p}
\eeq
or equivalently  in coordinate space basis \Eq{matrixlattice2} as
\beq
\widetilde \Phi = 
\sum_{{\bf n} \in \mathbb{Z}_{L'}^2} \Delta_{\bf n}^{'} \otimes 
\Phi_{\bf n}
\eqn{newbasis} 
\eeq
where   $\Delta_{\bf n}^{'}$ is the dual coordinate  basis. The 
$ J_{\bf p}^{\prime}$ and    $\Delta_{\bf n}^{'}$ are rank  ${L'^2k_1}$ 
matrices which provide the mapping between the $L' \times L'$ lattice and 
the matrix symbols. The $\Phi_{\bf n}$ are single valued,  
rank $k_0$ fields residing 
  on the dual noncommutative torus. 

Substituting 
the matrix symbols in the new basis  into \Eq{commutativematrix} 
and mapping  the trace as \footnote{ The traces are normalized such that 
$\Tr \,{\bf 1}$ gives the volume of discrete torus and 
$\tr \, {\bf  1}$ is the 
rank  of the gauge group. The other factors are there for 
the relative 
normalization. Namely, we request that the double trace of the 
corresponding identity matrices 
on both sides to be equal to each other.}           
\beq 
\Tr_{(L^2)}\, \tr_{(k)} \leftrightarrow  \Tr_{(L')^2k_1} \, \tr_{(k_0)} 
\, \frac{k}{k_0}\frac{Vol}{Vol'} \,,\qquad 
\eeq 
we  obtain the following action 
for the noncommutative $U(k_0)$ gauge theory: 
\begin{equation}
\begin{aligned}
S = \frac{k_0}{k}\frac{1}{g^2} \sum_{\bfn \in \mathbb Z_{L^{\prime}}^2}
 \tr_{(k_0)} 
\int d\theta\,\Bigl[& -\half \bfl_\bfn 
\star \partial_\theta  \bfl_\bfn -  \bfl_\bfn \star \left(\mybar
    z_{a,\bfn-\ah} \star 
    \bfz_{a,\bfn-\ah} - \bfz_{a,\bfn} \star \mybar
    z_{a,\bfn}\right) \\
&
-  \bfXi_{\bfn} \star ( \bfz_{1,\bfn} \star \bfz_{2,\bfn
  +{\bf \hat e_1} } - \bfz_{2,\bfn} \star \bfz_{1,\bfn + {\bf \hat e_2} })
\Bigr]\,.
\eqn{ncaction}
\end{aligned}
\end{equation} 

We have shown that at a nonperturbative level a $U(k)$ 
SYM  theory with a background 't Hooft flux (hence 
multivalued  fields) is  
equivalent to a purely noncommutative gauge theory $U(k_0)$ with periodic 
boundary conditions (hence single valued fields).  
The rank $k_0$ of the noncommutative theory is 
less than or equal to the rank of the commutative one.  The parameters of  the 
theory on the commutative lattice 
are the number of lattice sites $L \times L$, the coupling constant $g$, 
 the background 
magnetic flux $q=q_1k_0$, and the rank of the gauge group $k=k_0k_1$ 
where $k_0$ is greatest common divisor of $k$ and $q$.
 The parameters  in its 
noncommutative  equivalent are the number of lattice sites $L'\times L'$, 
the noncommutativity parameter $\theta'$, the  reduced rank of the 
gauge group $k_0$, and the coupling constant $g'$.  The relations among the 
parameters 
can be expressed as 
\begin{eqnarray}
&&
({L'})^2 = L^2 k_1^2 = L^2 \frac{k^2}{k_0^2} \nonumber \\ 
&& \theta'= \frac{b}{k_1}  \ , \qquad  bq_1 = 1 \,\, {\rm modulo}
 \,\,k_1 \nonumber \\  
&& g' = g ( \frac{k}{k_0})^{\frac {1}{2}}
\eqn{SL2Z1}
\end{eqnarray}
where $b$  is an integer. 
Notice that the total number of elementary  degrees of freedom are same in 
both theories.   
Even though the ranks of the gauge groups are  different, the volume of the 
dual torus is scaled in such a way that the total number of degrees of freedom 
is kept fixed.  
In the commutative case, we have $L^2$ unit cells (momentum modes) on the real 
lattice (momentum lattice) and each cell (mode)
has the matter content of the $\CN=(2,2)$ multiplet 
of $d=2$, $U(k)$ supersymmetric 
Yang-Mills theory. Hence there are  $ L^2  k^2$ degrees of freedom modulo 
matter content of  $\CN=(2,2)$ which is irrelevant for our discussion.  
In the noncommutative case, the lattice  has $(L')^2= L^2 k_1^2$ cells 
(momentum modes) and the 
gauge  group is $U(k_0)$ giving another factor of $k_0^2$. Thus, a total 
of  $L^2 k_1^2 k_0^2$ degrees of freedom  which is equal to the one 
in the commutative lattice. 

There are various cases that are particularly interesting:
 If $q$ and $k$ are coprime, the flux $q$ sector of  the $U(k)$
theory is equivalent to a noncommutative $U(1)$ theory. Notice that the 
$U(1)$ in $U(k)$ is free, 
noninteracting and hence decouples from the $U(k)$, leaving the $SU(k)$ 
sector. The  nonabelian features of the 
ordinary $SU(k)$  theory are captured by the spacetime noncommutativity of 
the  $U(1)$  theory.  
   
Let us assume 
$k$ to be a prime number. In that case, each of  the $k-1$ 
nontrivial topological 
sectors is mapped to a $U(1)$ theory with quantized units of 
noncommutativity 
$\theta'$. For the sector with flux $q$, the noncommutativity is 
$b \frac {1}{k}$  with $b.q=1$ modulo $k$. Thus $\theta'$ in dual $U(1)$
 theories is quantized in 
units of $\frac{1}{k}$.  
 
\noindent{\it Nontrivial topological sectors of noncommutative gauge theory}

Let us consider a  $U(k)$ supersymmetric gauge theory  with  a 
finite noncommutativity parameter  $\tilde \theta$ in the  
 background magnetic flux on an $L\times L$ discrete torus. 
Thus  the fields on the noncommutative lattice  are multivalued, and 
we would like to rewrite this theory as a purely noncommutative theory on 
some discrete torus with size   $L'\times L'$ with periodic boundary 
conditions.
  Following similar 
considerations as above, we find that the  dual  theory is related by an 
 $SL(2,\mathbb Z)$ transformation to the theory we started with. 
The  $SL(2,\mathbb Z)$ acts as 
\begin{eqnarray}
&&
({L'})^2 = L^2 (k_1 + q_1 \tilde \theta)^2   \nonumber \\ 
&& \theta'= \frac{b -a \tilde \theta }{k_1 + q_1 \tilde \theta }  \qquad  {\rm modulo}\,\,
 {\mathbb Z}  \nonumber \\ 
&& g' = g  (k_1 + q_1 \tilde \theta )^{\frac{1}{2}}
\eqn{SL2Z2}
\end{eqnarray}
where the matrix 
\begin{eqnarray}
\left( \begin{array}{cc}
        b & -a \\
        k_1 & q_1 
        \end{array} \right)
  \in SL(2,\mathbb Z).
\end{eqnarray}
Notice that for $\tilde \theta=0$,  the  transformations \Eq{SL2Z2} reproduce 
\Eq{SL2Z1}. From \Eq{mod2}, we know that   $\tilde \theta$  is not 
arbitrary, 
it satisfies  $L \tilde \theta \in 2 \mathbb Z$. Thus the size of the dual 
torus is 
$ {L'} = L k_1 + q_1 L \tilde \theta $ which is an integer as it should be. 
This result 
appeared in string and M-theory  as part of the T duality group 
\cite{Connes:1997cr, Seiberg:1999vs, Brace:1998ku, 
 Pioline:1999xg, Schwarz:1998qj, Hofman:1998iy} and in  the context 
of fully regularized pure Yang-Mills in \cite{Ambjorn:2000cs}. In terms of 
dual gravity picture, it is discussed in \cite{Cai:2000hn, Cai:2000yk}. 
 Hence, we will 
not discuss it here in detail. 

However, let us reconsider the $U(k)$ gauge theory with  a prime $k$
and its  nontrivial topological 
sectors. These sectors have  purely noncommutative $U(1)$ duals. If the $q$
flux sector is equivalent to a theory with noncommutativity $\tilde \theta$, 
then the $q'=k-q$ flux  sector is equivalent to a theory with noncommutativity 
$\theta'$ such that $\tilde \theta + \theta'=1$. The $U(1)$ theory with 
noncommutativity $\theta'$ is dual  to  a $U(1)$ with noncommutativity  
$\tilde \theta$ by an $SL(2, \mathbb Z)$ transformation, which is given by 
$S=\left( \begin{array}{cc}
        1 & -1 \\
        1 & 0 
        \end{array} \right)$. 
   
\section{Noncommutative  $\CN= (4,4)$  in  $d=2$ }
The prescription we described in section \ref{sec:reg} can be applied almost 
verbatim  to generate a regularization for noncommutative $\CN= (4,4)$  
extended SYM theories in $d=2$ dimension. 

The action of the parent   matrix theory is a $\CQ=2$ the deformation of 
dimensional  reduction of $\CN=1$  SYM theory with $U(N_c)$ gauge group 
from $d=6$ 
Euclidean dimensions to zero dimension.  The undeformed zero dimensional 
model possess a $\CQ=8$ supersymmetry and  
a global $SO(6) \times SU(2)$ R-symmetry group. The $SO(6)$ is inherited 
from the Lorentz symmetry  and $SU(2)$  is the R-symmetry group prior to 
reduction. The matter  content of the theory is a six component 
gauge vector potential $\{z_a, \mybar z_a\},\, a= 1, \dots 3$ expressed as 
complex matrices   and an eight 
component spinor $\{\xi_a, \psi_a, \chi, \lambda\}, \, a= 1, \dots 3$.
  There are also auxiliary fields $\{d, G, \mybar G\}$ 
introduced for off-shell supersymmetry.   
 
We begin with a little digression to  $\CQ=2$  supersymmetry and its 
supermultiplets in terms of which  
the deformed matrix model action 
can be written  in manifestly $\CQ=2$ form \cite{Cohen:2003qw}. We will 
work with a superspace with two independent Grassmann valued coordinates 
$(\theta, \mybar \theta )$.  
There are two important nilpotent symmetries, $Q$ and $\mybar Q$.  
The action of  supersymmetry $Q$ on component fields is  
\begin{eqnarray}
\begin{array}{lllllll}
%
Q z_i &=& \sqrt{2}\,\psi_i & \qquad & Q \mybar z_i
&=& 0  \\
Q \psi_i&=& 0 &\qquad&
Q \xi_i&=& -2  \,[\mybar z_3,\,\mybar z_i]\\  &&&&&&\\
Q z_3 &=& \sqrt{2} \psi_3 &\qquad& Q \lambda &=&- \,\left( id
+ [\mybar  z_3,\,z_3]\right)\\
Q \psi_3&=&0
&\qquad\qquad  &
Q (i d) &=& - \sqrt 2 [\mybar  z_3,\,\psi_3] \\  &&&&&&\\
Q \xi_3 &=&\sqrt 2 G 
 &\qquad& 
  Q \chi&=& 0 \\
Q G &=& 0 
 &\qquad& 
Q \mybar G &=&  -2[\mybar z_3, \chi] \ ,
\end{array}
\eqn{Qaction}
\end{eqnarray}
and $Q \mybar z_3=0$ where  $i=1,2$.  The other supersymmetry 
$\mybar Q$ acts as 
\begin{eqnarray}
\begin{array}{lllllll}
\mybar Q z_i &=& 0  & \qquad & \mybar Q \mybar z_i
&=&  \,\sqrt{2}\,\,\xi_i \\
\mybar Q \psi_i&=& -2 \,[\mybar z_3, z_i]  &\qquad&
\mybar Q \xi_i&=& 0 
 \\ 
 &&&&&&\\
\mybar Q z_3 &=& \sqrt{2} \lambda 
&\qquad&  
\mybar Q \psi_3&=& \,\left( id 
 - [\mybar z_3,\,z_3]\right)
\\
\mybar Q \lambda&=& 0
&\qquad\qquad  &
\mybar Q (i d)  &=&  \sqrt 2 [\mybar z_3, \lambda] 
 \\  
 &&&&&&\\
\mybar Q\xi_3 &=& 0 
 &\qquad& 
\mybar Q \chi&=& \sqrt 2 \, \mybar G \\
\mybar Q G &=&  -2 [ \mybar z_3, \xi_3]
 &\qquad& 
\mybar Q \, \mybar G &=& 0  \ ,
\end{array}
\eqn{Qbar}
\end{eqnarray}
and $\mybar Q \mybar z_3=0$ where $i=1,2$. Since
 ${\mybar z_3}$ is in the intersection of the kernel of $Q$ and 
$\mybar Q$, it is a supersymmetry singlet.  
In order to construct the supersymmetry multiplets, we introduce 
two differential operators  $Q$ and $\mybar Q$, realized 
in the superspace as  
\begin{equation}
Q =\frac{\partial\  }{\partial \theta} + \sqrt{2}\, 
\mybar\theta[\mybar z_3,\,\cdot \ ] \ ,\qquad
\mybar Q = \frac{\partial\  }{\partial \mybar\theta} +   
\sqrt{2}\, \theta[\mybar z_3,\,\cdot \ ] \ .
\end{equation}
with the  algebra \qquad 
\beq
 Q^2 = \mybar Q^2 =0, \,\;\; \{Q, \mybar Q\}\,\cdot= 2 
\sqrt 2 [\mybar z_3,\,\cdot ]
\eqn{sual}
\eeq  
Notice that the supersymmetry transformations \Eq{Qaction} and 
\Eq{Qbar}
are satisfied off-shell.\footnote{The supersymmetry algebra is the same 
as  \Eq{sual}, the algebra of differential operators, 
except that the sign of the right hand side of the 
anticommutator  $\{Q, \mybar Q\}$  is switched. For an explanation, 
see \cite{Wess:1992cp}, page 25-26.}
We can also define supersymmetric derivatives which anticommute with
the $Q$'s:
\begin{equation}
{\CD}=\frac{\partial\  }{\partial \theta} - \sqrt{2}\, \mybar\theta\,
[\mybar z_3,\,\cdot \ ] \ ,\qquad
{\mybar\CD} =\frac{\partial\  }{\partial \mybar\theta} -  \sqrt{2}\, \theta\,
[\mybar z_3,\,\cdot \ ] \ .
\end{equation}
The supersymmetric derivatives may be used in constructing the action.  

The transformations in the first two lines of  \Eq{Qaction} and 
\Eq{Qbar}
are compatible with    bosonic chiral and antichiral superfields 
with a superspace expansion 
\begin{eqnarray}
&&{\bfz_i} = z_i +\sqrt{2}\,\theta \psi_i -\sqrt{2}\, \theta\mybar\theta
[\mybar z_3,z_i]\ ,
\nonumber \\
&&\mybar {\bfz}_i = \mybar z_i
+\sqrt{2}\,\mybar\theta \, \xi_i +\sqrt{2}\,
 \theta\mybar\theta [\mybar z_3,\mybar z_i] \ ,
\end{eqnarray}
satisfying the chiral and antichiral constraints  ${\mybar\CD}{\bfz_i}=0$, 
and  ${\CD}\mybar {\bfz}_i= 0$. 
The last two lines of the transformations  \Eq{Qaction} and \Eq{Qbar} 
are in accord with   fermi multiplets with superspace expansions 
\begin{eqnarray}
&&{\mathbf \Xi} = \xi_3 +\sqrt{2}\, \theta G  -\sqrt{2}\, \theta \mybar\theta 
[\mybar z_3,\xi_3]\ ,\quad 
 \nonumber \\
&& \mybar{\mathbf \Xi} = \chi +\sqrt{2}\,\mybar\theta \mybar 
G  +\sqrt{2}\,\, \theta \mybar\theta 
[\mybar z_3,\chi]\ ,
\end{eqnarray}
satisfying chiral and antichiral  constraints  
${\mybar\CD}{\bf \Xi}=0$  and  ${\CD} \mybar {\bf \Xi}=0$. 
The transformations  in the middle two lines of 
\Eq{Qaction} and \Eq{Qbar} combine to a vector multiplet \footnote{ 
The vector multiplet in Minkowski space satisfies a reality 
(Hermiticity) condition. The 
Euclidean space vector multiplet should be interpreted as   satisfying the 
reality condition after analytical continuation to Minkowski space. The 
complex matrices $z_3$ and $\mybar z_3$ turn into Hermitian matrices upon
analytic continuation. (For example, compare with the dimensional 
reduction of $(0,2)$ multiplets  appearing 
in \cite{Witten:1993yc} to zero dimension)  
In this 
regard, $\lambda$  and $\psi_3$
which are independent Grassmann variables 
in Euclidean space turn into complex conjugate variables in Minkowski space. 
(similarly for fermionic components of bosonic 
chiral and antichiral multiplets and highest components of  
fermi multiplets.)} 
\begin{equation}
{\bfs} = z_3 + \sqrt{2}\,\theta \psi_3 + \sqrt{2}\, \mybar\theta \lambda
+\sqrt{2}\,\theta \mybar\theta (id)\ .
\end{equation}
From the vector superfield we can create the chiral and anti-chiral
superfields which will appear in the action:
\begin{equation}
\begin{aligned}
{\mathbf \Upsilon}&= \frac{\mybar{{\CD}}{\bfs}}{\sqrt2}= \lambda - 
\theta (+[\mybar{z}_3, z_3] +i d)
-\sqrt2 \theta \mybar{\theta}[\mybar{z}_3, \lambda]   , \\
 \mybar {\mathbf\Upsilon}&= \frac{{\CD}{\bfs}}{\sqrt2}= \psi_3 + 
\mybar{\theta}(
 -[\mybar{z}_3, z_3] +i d) 
+ \sqrt2 \theta \mybar{\theta}[\mybar{z}_3,\psi_3] \ .
\end{aligned}\end{equation}
with ${\mybar\CD}{\bf \Upsilon}=0$  and  ${\CD} \mybar {\bf \Upsilon}=0$.

Note that the $\theta$ ($\mybar
\theta$) components of (anti-)chiral superfields, and the
$\mybar\theta\theta$ components of a general  superfield, transform under
supersymmetry into a commutator;  therefore, the trace of such terms
are supersymmetric invariants and are suitable for construction of the
action. In terms of these superfields, the action of the $\CQ=2$ deformed  
matrix model  may be written as
\begin{equation}\begin{aligned}
S &= \frac{1}{g^2} \int d\theta d\mybar\theta\, \Tr \left(
    \frac{1}{2}\mybar{\mathbf\Upsilon}{\mathbf\Upsilon} + \frac{1}{\sqrt2} 
{\mybar\bfz}_i 
[{\bfs},{\bfz_i}]
+ \frac{1}{2}\mybar{\mathbf\Xi} 
{\mathbf \Xi}\right)
\\ &\qquad +\int d\theta \, \Tr\left({\mathbf \Xi}\,[{\bfz_1},{\bfz_2
  }]_{\zeta} \right)
 -\int d\mybar\theta \, \Tr\left(\mybar{\mathbf\Xi}\,
[\mybar {\bfz}_2, 
{\mybar\bfz}_1]_{\zeta^{*}}\right)
\ ,
\end{aligned}
\eqn{twisted2}
\end{equation} 
where $\zeta$ can be regarded as a deformation of the superpotential 
and is set 
to a specific value following the discussion in  \ref{subsec:moral}.  
Similar to  section \ref{sec:reg}, we can orbifold the $\CQ=2$  deformed 
  matrix model 
to obtain the lattice action for $\CN=(4,4)$  noncommutative 
SYM theory where the lattice 
possess $\CQ=2$ exact supersymmetries. To orbifold, we use conveniently 
chosen {\bf r}-charges, given in Table \ref{tab:tab2nc}.   
\setlength{\extrarowheight}{5pt}
\begin{table}[t]
\centerline{
\begin{tabular}
{|c||c|c|c|c||c|c|c|}
\hline
&$  \bfz_1 $&$\mybar  \bfz_1 $&$ \bfz_2$&$\mybar  \bfz_2$&$ {\bf S},
{\bf \Upsilon}, {\mybar {\bf \Upsilon}} $& $ \bfXi $ & $\mybar  \bfXi $
\\ \hline
$ r_1 $&$ +1 $&$ -1 $&$ \,\ 0 $&$ \,\ 0 $&$ \,\ 0 $&$ -1 $&$ +1 $
\\
$ r_2 $&$\,\ 0 $&$ \,\ 0 $&$ +1 $&$ -1 $&$ \,\ 0 $&$ -1 $&$ +1 $
\\ \hline
 \end{tabular}
}
\caption{\sl The  $r_{1,2}$ charges  of the  fields of the $\CQ=2$  
deformed matrix theory
which  define the orbifold projection.\label{tab:tab2nc}}
\end{table}

The  lattice  action for noncommutative $\CN= (4,4)$ SYM theory with 
gauge group $U(k)$ is written with 
manifestly $\CQ=2$ superfields as  
\begin{equation}\begin{aligned}
S= \frac{1}{g^2_{nc}}\sum_{\bfn \in \mathbb Z_L^2}
\tr_{(k)} 
&\left[\int d\theta d\mybar\theta\,  \left(
    \frac{1}{2}\mybar{\mathbf\Upsilon}_{\bfn}\star {\mathbf\Upsilon}_{\bfn} 
+\frac{1}{\sqrt{2}}{\bfs}_{\bfn} \star ( {\bfz}_{a,\bfn} \star 
{{\mybar\bfz}}_{a,\bfn} -  {{\mybar\bfz}}_{a,\bfn-
    \ah} \star {\bfz}_{a,\bfn-\ah})
+\frac{1}{2}\mybar{\mathbf\Xi}_{\bfn} \star 
{\mathbf \Xi}_{\bfn}  \right) \right.\\ 
&\quad +  \left.
\int d\theta\,\Bigl(
 {\mathbf \Xi}_{\bfn }\, \star ( {\bfz}_{1,\bfn} \star 
{\bfz}_{2,\bfn + {\bf \hat e_1} }  - {\bfz}_{2,\bfn } \star {\bfz}_{1,\bfn + 
{\bf \hat e_2}  })
\Bigr) + a.h.   
\right]
\end{aligned}
\eqn{lact2nc}
\end{equation}
where $a.h.$ stands for the anti-holomorphic part of the superpotential. 
The chiral and antichiral superfields forming the lattice 
action satisfy  highly nonlocal constraints, such as  
\beq 
\mybar \CD {\bfz}_{1, \bfn}= \frac{\partial\  }{\partial \theta}
{\bfz}_{1, \bfn} - 
\sqrt{2}\, \mybar\theta\, ( \mybar z_{3,\bfn}\star  {\bfz}_{1, \bfn} 
-  {\bfz}_{1, \bfn} \star   \mybar z_{3,\bfn +{\bf \hat e_1} }) \; = 0 
\eeq
which implies some  components of the superfields are nonlocal. The explicit 
form of the lattice superfields can be easily obtained by using 
\Eq{matrixlattice}. The discussions of the continuum limits, 
the reason for deformation 
of the superpotential and  discussion of dualities 
 are the  same as $\CN=(2,2)$ SYM theory and will not be repeated here.  

\section{Noncommutative $\CN=(8,8)$ in $d=2$}
\label{sec:nc8,8}
The last target theory for which we construct a lattice  regularization 
is maximally supersymmetric $\CN=(8,8)$ noncommutative 
SYM  theory in two dimensions. 
The action of the deformed matrix theory is
a $\CQ=4$  deformation of the dimensional reduction of $\CN=1$ SYM theory from 
$d=10$ dimensions to $d=0$ dimension. 
 The $\CQ=4$ supersymmetry is the $\CN=1$ 
supersymmetry of the $d=4$ dimensions, hence the multiplets are  familiar 
 from  supersymmetry in four dimensions (see for example \cite{Wess:1992cp}).
We will denote the three chiral (antichiral) multiplets with 
$\bfz_a \, ( \mybar  \bfz_a) $, the vector multiplet with ${\bf V}$ and 
 the chiral (antichiral) fermi  multiplet with   ${\bf W},
\mybar {\bf W} $.
The holomorphic superpotential of the $\CQ=4$  deformed matrix model is 
\beq
 \Tr \bigl( {\bfz}_3 
[\bfz_1, \bfz_2]_\zeta   \bigr)
\eqn{deform}
\eeq
where $\zeta$ is the deformation parameter.\footnote{In the four 
dimensional  counterpart of this  matrix model,  
the deformation  of the superpotential \Eq{deform} is  an
exactly marginal  
deformation and it was examined in  \cite{Leigh:1995ep}.}  
For $\zeta=1$, the matrix model possess all $\CQ=16$ supersymmetries and a 
global $SO(10)$ R-symmetry.  
  
The orbifold projection of the $\CQ=4$ deformed matrix model  
can be performed  by using the ${\bf r}$-charges under the $U(1)\times U(1)$ 
subgroup 
of the  global  symmetry group of the  matrix model 
given in  Table \ref{tab:tab3nc}. 
\setlength{\extrarowheight}{5pt}
\begin{table}[t]
\centerline{
\begin{tabular}
{|c||c|c|c|c|}
\hline
&$  \bfz_1 $&$  \bfz_2 $&$ \bfz_3$&${\bf V} $
\\ \hline
$ r_1 $&$ +1 $&$ 0 $&$ \,\ -1 $&$ \,\ 0 $
\\
$ r_2 $&$\,\ 0 $&$ \,\ +1 $&$ -1 $&$ 0 $
\\ \hline
 \end{tabular}
}
\caption{\sl The  $r_{1,2}$ charges  of the  fields of the $\CQ=4$  
deformed matrix theory
which  define the orbifold projection.\label{tab:tab3nc} The charges of the 
antichiral multiplets are negative of their chiral counterparts.  }
\end{table}
 
The lattice action preserves four exact  supersymmetries. For each unit cell, 
there are three chiral multiplets and three antichiral ones residing on the 
links. The vector multiplet resides on the sites.
The lattice action for noncommutative $\CN=(8,8)$  theory may be written 
in manifestly $\CQ=4$ supersymmetric form as   
\begin{equation} 
\begin{aligned}
S=
\frac{1}{g_{nc}^2}\sum_{\bfn \in \mathbb Z_L^2} \Tr\biggl[&
 \int  d^2\theta \; d^2 \mybar \theta \;\; \mybar {\bfz}_{a,\bfn} \star  
e^{2{\bf V}_{\bfn}} \star  {\bfz}_{a, \bfn} \star  e^{-2{\bf V}_{\bfn+ \ah}} + 
\; \frac{1}{4}
\int d^2  \theta \; 
{\bf W}^{\alpha}_{\bfn} \star {\bf W}_{\alpha, \bfn} +  a.h. 
\\ 
&+ \sqrt{2}  \int d^2 \theta  \,\, {\bfz}_{3, \bfn}\star   
( \bfz_{1, \bfn } \star \bfz_{2, \bfn + \hat{\bf e}_1 } - \bfz_{2, \bfn } 
\star 
\bfz_{1, \bfn +\hat{\bf e}_2 } ) + a.h. 
\biggr]
\end{aligned}
\end{equation}
where $a.h.$ stands for the antiholomorphic superpotential. The first 
term in the 
action is the standard K{\"a}hler term, and it also reflects the star gauge 
transformation  properties of the chiral link fields. The second term 
 must be  interpreted as an holomorphic 
superpotential of a chiral fermi multiplet ${\bf W}_{\alpha}$ but not as a 
field strength superfield (similarly for its antiholomorphic counterpart).  
The last  term, which is a   
superpotential  gives rise to gauge kinetic terms ( as well as other terms) 
in the 
continuum target theory. 
We would like to emphasize that the gauge boson of the continuum theory 
does not reside in the site multiplet ${\bf V}$, 
but on the chiral link multiplets. Similarly, the scalars of the continuum 
theory (there are eight of them) live both on the chiral link multiplets 
and on the site multiplet ${\bf V}$. For more details of the commutative 
counterpart of this lattice, see section 5 of Ref.\cite{Kaplan:2005ta}.

\section{Discussion and outlook}
In this article, we have examined nonperturbative aspects of extended SYM 
theories in $d=2$ Euclidean dimensions. 
We first constructed a full star gauge invariant and manifestly supersymmetric 
non-perturbative regulator for noncommutative supersymmetric gauge theories.  
 
Then we observed that SYM theories may be classified  as 
{\it (i)}  noncommutative, with no background  flux,   
{\it (ii)} noncommutative, with a background  flux, 
{\it (iii)} commutative, with a background  flux, 
{\it (iv)}  commutative, with no background flux.  
The theories in the first three class are related to each other 
by an $SL(2, \mathbb Z)$ transformation. In particular, we have explicitly 
shown that the 
nontrivial topological sectors of SYM theories may be rewritten as 
purely noncommutative field theories.   
The commutative sector with no  background flux is not related by an 
$SL(2,\mathbb Z )$  to other sectors.  

The noncommutative SYM theory with rank $k_0$ and noncommutativity 
parameter $\theta'$ 
given in \Eq{ncaction}  
may be  regarded as regulated via a commutative field theory 
\Eq{commutative} with rank $k$ and background  flux $q$ obeying twisted 
boundary conditions \Eq{twisted}. This brings us to the question of 
renormalizability and the continuum of quantum theory.
In \cite{Cohen:2003xe, Cohen:2003qw}, it has been shown that  the target 
continuum  theory of the commutative lattice action is obtained  
without any fine tuning  owing to the  superrenormalizability of the 
theory (modulo an uninteresting Fayet-Illiopoulos term).
Also, it has been argued that the fluctuation of the  modulus in the  
quantum theory does not destroy the lattice interpretation as long as the     
continuum limit is taken such that 
${\rma}^2g_2^2{\rm ln}( L\rma / {\rma} ) \rightarrow 0$.
(For a fuller explanation, see  \cite{Cohen:2003xe, Cohen:2003qw}.)
We believe that these two facts imply  the corresponding 
quantum noncommutative theory is a renormalizable theory 
with a sensible continuum limit. 

The observables in the zero momentum sector of the commutative gauge theory  
with flux can be mapped to the star gauge invariant observables which also 
carry zero momentum in  the noncommutative theory.
 In the case of pure 
Yang-Mills theory, the mapping of the zero momentum sector along with the 
translational invariance of the two theories 
may be used to obtain the loop equations and to show 
  the relation among at least a subset of observables 
in two  theories along the lines of \cite{Kovtun:2003hr}  
in the large $N_c$  limit. 

On the other hand, the noncommutative SYM theories we have examined have
  no local 
gauge invariant observables. The noncommutativity of spacetime puts 
a stringent  bound on the size of a gauge invariant  excitation. 
In particular, if we want to minimize smearing 
of an excitation in both directions, the bound 
$\Delta x_1 \Delta x_2 \gtrsim \Theta_{12}$ 
requires that 
$\Delta x_a \sim \sqrt \Theta $ assuming $\Theta$ is larger then the 
other length scale in the problem $\frac{1}{g_2^2}$
. Thus the
best localization that can be achieved in the theory is (in terms of lattice 
parameters) $L {\rma} \sqrt{\theta'}$ where  $L^2$ is the number of sites and 
${\rma}$ is lattice spacing and $\theta' \in [0,1)$ is dimensionless 
noncommutativity  on the lattice.   We see that the dimensionless ratio 
of the size of an excitation to the size of the box is 
\beq
\frac{\Delta x_a}{L\rma} \sim \sqrt {\theta'} \, .  
\eeq
  The existence of the  Seiberg-Witten map 
\cite{Seiberg:1999vs} from commutative 
gauge fields to noncommutative gauge fields 
suggests that the local gauge 
invariant   observables in commutative theory may be used to construct 
the nonlocal gauge invariant counterparts in the noncommutative theory 
 where the latter are smeared over a region of size
 $\sqrt {\Theta}$ \cite{Minwalla:1999px} consistent with what we argued 
above. For observables in noncommutative field theories,  see  
\cite{Ishibashi:1999hs}.

One other issue we did not discuss is 
 the irrational dimensionless noncommutativity 
parameters, which requires taking the infinite $N_c$ limit  
\cite{Ambjorn:2000cs} of ordinary  lattice  SYM theory.
The large but finite $N_c$ gauge 
theories  have flux sectors which are equivalent to a finite rank  
gauge theories with rational noncommutativity parameters.

The study of three  and four dimensional   noncommutative 
extended SYM theories is left for the future work. 
Even though generalization of the results of this paper is straightforward, 
the theories in  three and four dimensions are richer both 
topologically and  dynamically.   For example,  in four dimensions,
the theory has sectors with  both electric and magnetic 
fluxes as well as instantons.    
Following \cite{'tHooft:1979uj} and the arguments given in section 
\ref{sec:Tduality}, 
the dualities between different noncommutative theories can be established 
rigorously at the non-perturbative level.    

\acknowledgments We would like to thank D. B. Kaplan, 
A. Karch, P. Kovtun, 
L. G. Yaffe  for numerous discussions,  A. Connes for his delightful 
lectures on 
noncommutative geometry at Seattle and conversations, M. Douglas for 
conversations,  
 A. B. Clark for 
reading the manuscript and suggestions. 
This work is  supported in part by DOE grant DE-FGO2-00ER41132,

\newpage
\sloppy

\end {document}